\documentclass[journal,12pt,onecolumn,draftclsnofoot]{IEEEtran}
\IEEEoverridecommandlockouts

\usepackage{cite}
\usepackage{amsmath,amssymb,amsfonts}
\usepackage{algorithm}
\usepackage{algorithmic}
\usepackage{graphicx,psfrag}
\usepackage{graphicx}
\usepackage{textcomp}
\usepackage{xcolor}
\usepackage{amsthm}
\usepackage[justification=centering]{caption}
\usepackage{overpic}
\usepackage{multirow}
\usepackage{fancyhdr}
\usepackage{lmodern}

\title{Decision Feedback Differential Detection for Reconfigurable Intelligent Surfaces}
\author{Jiawei Qiu \footnote{Jiawei Qiu was with the Department of Electrical and Computer Eng. McGill University, Montreal QC Canada. He is now with INRS Telecomunications, University of Quebec, Montreal QC Canada.}
 and
 Harry Leib \footnote{Harry Leib is with the Department of Electrical and Computer Eng., McGill University, Montreal, QC Canada, Email: harry.leib@mcgill.ca}	}

\begin{document}
	\maketitle
	
	\begin{abstract}
		This work considers a Differential Reflecting Modulation (DRM) scheme for Reconfigurable Intelligent Surfaces (RIS) not requiring channel state information (CSI). When operating over time-varying fading channels, such schemes  with Conventional Differential Demodulation (CDD) receivers experience high error floors and performance degradation. To address these issues, we propose a Decision Feedback Differential Detection (DFDD) technique for DRM. We explore the application of DFDD for RIS DRM and conduct extensive Monte-Carlo simulations to analyze performance. Results demonstrate the viability of our DFDD technique across various RIS scenarios and highlight the importance of proper parameter selection to achieve good performance. The DFDD scheme is also compared with  uncoded  and Differential Space-Time Modulation (DSTM) coded DRM  using CDD based receivers. We observe that at low SNR, the DFDD scheme performs almost as well as the DRM with CDD scheme, but worse than the DSTM coded DRM. As the SNR increases however, both CDD-detected systems encounter high error floors while the error rate of DFDD based scheme continues to improve until it reaches a relatively low error floor.
It is shown that the chief merits of employing DFDD receivers in such RIS systems is the low error floors they provide over time varying fading channels, albeit at expense of a small increased complexity.
	\end{abstract}
	
	\section{Introduction}
	The  Reconfigurable Intelligent Surface (RIS) technology, also referred to as Intelligent Reflecting Surface (IRS), holds significant potential for facilitating an advantageous wireless communication environment~\cite{umer2025recon, shi2024ris, fu2025multi, wu2026int,
  RISee, smart}. An RIS system consists of a large number of passive reflecting units, each capable of altering the incident signals in terms of phase and amplitude. These units are made from electromagnetic material, with size and spacing much smaller than the wavelength. Unlike relays, RIS can be constructed with minimal hardware complexity and cost, employing low-power and low-complexity electronic circuits~\cite{RISvsRelay}. The primary characteristic of RIS is its reconfigurability after deployment in a wireless environment, allowing for a programmable and controllable communication environment~\cite{SmartRIS}.
	
	Recently, there has been a notable  interest in research on RIS communication techniques. A substantial volume of literature has been dedicated to exploring RIS  imodulation techniques~\cite{yue2026hybrid, yue2026ris, zhang2026ris,basar2020,basar2019wireless,RM}, and channel state information (CSI)\cite{alwazani2020, wang2025low, de2025channel, zhuo-liuchang, zheng2025mutual}. 
Alwazani \textit{et al} in~\cite{alwazani2020} introduced an RIS-assisted multi-user multiple-input single-output (MISO) communication system under imperfect CSI, demonstrating the efficiency of this scheme and its high sensitivity to the quality of the estimates. However, to the best of our knowledge, most  schemes assume perfect CSI for detection~\cite{chen2026environment, gao-lu, perfectCSI3,perfectCSI4}. Channel estimation in an RIS-based communication system has to consider the transmitter-RIS, RIS-receiver, and direct transmitter-receiver links. Such estimations schemes are complicated due to the absence of baseband signal processing capabilities in RIS units. While efforts have been made to address the challenges of channel estimation in RIS systems ~\cite{imperfectCSI1,imperfectCSI2,alwazani2020, gao-lu}, CSI acquisition  remains a significant and challenging problem, making  tranmission techniques not requiring CSI attractive..
	
	Noncoherent detection is a demodulation technique that does not require CSI, thereby reducing the complexity of communication systems. Moreover, noncoherent demodulation could be more reliable than coherent demodulation, particularly at low SNR, due to the latter's reliance on CSI which presents challenges to acquire \cite{noncobetter}. As a result, RIS-assisted systems employing noncoherent detection techniques have attracted considerable interest \cite{mishra2025ser, cai2025design, nonco1,nonco2,nonco3,nonco4,nonco5}. In \cite{nonco1}, Cai \textit{et al} present a noncoherent RIS-assisted joint index keying $M$-ary differential chaos shift keying scheme. This system enables the receiver to employ noncoherent correlation without the need for CSI, thus reducing system complexity. Simulation results indicate that this technique outperforms existing systems. In \cite{nonco3}, an RIS-aided differential chaos shift keying technique is proposed to overcome the limited BER performance gains observed in noncoherent chaotic communications. By deploying RIS near the transmitter antenna, the scheme enhances the SNR of the received signal, significantly improving error performance.

	Despite of many advantages of noncoherent demodulation, \cite{DRM} noted that conventional differential detection (CDD) for RIS, which a common form of noncoherent detection,  results in performance degradation. For scalar communication systems, more advanced noncoherent detection techniques have been introduced to improve performance.. The concept of Decision Feedback Differential Detection (DFDD)  was first proposed by \cite{WHequation} and analzed  in \cite{firstDFDD}. The DFDD technique  utilizes feedback from previously detected  symbols to establish a demodulation reference, leading to a posssible  iimprovement in  BER performance over CDD that employs only the previous symbol as a reference.  Demodulators based on DFDD have been initially considered for AWGN channels in \cite{WHequation} and later extended to fading channels in \cite{DFDDflatfading,linearpredictDFDD}. In \cite{DFDDflatfading}, it was concluded that DFDD employing finite-order feedback filters can significantly reduce the error floor compared to CDD. Due to their effectiveness in scalar transmission systems,  DFDD techniques have triggered research into corresponding detection methods for multi-antenna scenarios. In \cite{MSDD}, DFDD based structures were applied to unitary-matrix signaling-based DSTM over Multiple-Input Multiple-Output (MIMO) fading channels.
	
	Our paper considers, a DRM scheme from \cite{DRM} for RIS systems.  Different from \cite{nonco1,nonco2,nonco3,nonco4,nonco5}, this DRM scheme operates on $M$-ary PSK. Information bits are mapped to both the activation permutations order of the reflecting patterns and the phases of transmitted signals, which facilitate CDD based demodulation, avoiding channel estimation. However, CDD may result in performance loss and error floor over time-varying fading channels for DRM. This motivates  to consider a DFDD technique for DRM RIS systems  operating over time-varying Rayleigh fading channels Our work demonstrates that the use of DFDD techniques for such RIS systems lowers significantly the error floors when compared to CDD.
	
	The rest of this paper is organized as follows. In section \ref{section:2}, we introduce the system model of the DRM RIS system. The fundamentals of  DFDD demodulation, its extension to DRM and performance issues are presented in Section \ref{section:3}   Section \ref{section:4} offers a performance comparison between  DFDD DRM  against  uncoded DRM and  DRM-DSTM coded schemes employing CDD, since the latter  has also error floor reduction capabilities. In our paper we also evaluate the associated  computational complexities. Conclusions are drawn in Section V, and Appendix A presents the details of the DFDD  prediction coefficients derivation.

	\section{System model and fundamentals of  DFDD-detected DRM} \label{section:2}
	
	We consider a model for an RIS system employing Differential Reflecting Modulation (DRM) as in \cite[Figure 1]{qiu2026space}. Let $\mathbf{X}[t]\in\mathbb{C}^{K \times K}$ 
	be the information carrying matrix   $\mathbf{X}[t]=\mathbf{Z}[t]\mathbf{S}[t]$ where $\mathbf{Z}[t]$ is a
	 $K \times K$ permutation matrix, and $\mathbf{S}[t]$ is a $K \times K$ diagonal matrix of MPSK symbols \cite{qiu2026space}. The number of information bits that 
	 $\mathbf{X}[t]$ carries is shown in \cite{qiu2026space} to be
	 \begin{equation}
	 r=\lfloor \log_2 K! \rfloor + K \log_2 M
	 \label{eq:r-bits}
	 \end{equation}
	 where $M$ is the MPSK constellation order.
	Then matrix differential encoding is performed
	\begin{equation}
	\mathbf{V}[t]=\mathbf{V}[t-1]\mathbf{X}[t], \label{eq:diffencode}
	\end{equation}
	where $\mathbf{V}[t-1]$ is generated in the previous block. The first block in a frame does not convey information, thus $\mathbf{V}[0]$ is  set as an identity matrix.
	
	\begin{figure}[htbp]
		\centering
		\includegraphics[width=0.4\linewidth]{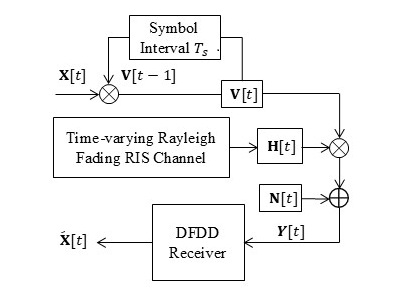}
		\caption{DFDD-detected DRM encoding structure}
		\label{DFDDsystem}
	\end{figure}
The equivalent system to  \cite[Figure 1]{qiu2026space} we consider in our present paper  aiming  at deriving a DFDD receiver for a DRS RIS system  can be represented by:
	\begin{align}
		\boldsymbol{Y}[t]=\mathbf{H}[t]\mathbf{V}[t]+\mathbf{N}[t], \label{eq:DFDDreceive}
	\end{align}
and it is illustrated in Fig.\ref{DFDDsystem}. The equivalent channel matrix \cite{qiu2026space} is $ \mathbf{H}[t]$ which is determined by the channels from the transmitter to RIS, from RIS to receiver including RIS reflecting pattern selection,  and also by the direct channel from the transmitter to receiver.  The additive noise matrix is $\mathbf{N}[t]\in\mathbb{C}^{N_r\times K}$.
The channels are represented as  zero-mean complex Gaussian processes (time varying Rayleigh fading based on Jakes  model). of unity normalized power, and spatially uncorrelated \cite{OnDFDD}.
The autocorrelation function (ACF) of the symbol-rate-sampled fading process is given by $ J_0(2\pi f_DmT_s)$
where $J_0(\cdot)$ denotes the zeroth-order Bessel function of the first kind, and $f_DT_s$ represents the normalized Doppler frequency \cite{OnDFDD}.  At values of $f_DT_s$ used in this work the fading process essentially remains constant over a time  interval of duration $T_s$, hence we assume the absence of intersymbol interference \cite{linearpredictDFDD}. This assumption holds more strongly when  $T_s\leq \frac{1}{f_D}$. In our simulations, we set the normalized Doppler frequency $f_DT_s\leq0.03$.
The additive  noise $\mathbf{N}[t]$ follows an uncorrelated complex Gaussian distribution with zero mean and covariance $\sigma^2\mathbf{I}_{N_r}$. The reflecting pattern selection procedure is detailed in \cite{qiu2026space}. While in
\cite{qiu2026space} the receiver is based on CDD,  in our present paper we consider DFDD for DRM RIS systems.
	
	The core concept of the DFDD technique involves using previously decoded symbols to feed back and used to establish a demodulation reference assisting the detection of a subsequent symbol.	
	Following the derivation of the DFDD receiver for scalar systems as presented in \cite{DFDDflatfading,linearpredictDFDD}, subsequent research has focused on applying DFDD detection techniques to multi-antenna systems. Notably, two similar formulations for the Differential Space-Time Modulation (DSTM) scheme, which employs decision-feedback differential demodulation, were introduced in \cite{WHinMatrix1} and \cite{WHinMatrix2}. In \cite{WHinMatrix1}, it is highlighted that for multi-antenna systems, the scalar predictor coefficients are derived based on the assumption of spatially uncorrelated fading. This assumption ensures that the fading processes are uncorrelated across different spatial paths. In DFDD receivers, the time correlation, however, plays a crucial role. The faster is the fading process, the more performance loss can be expected. In our work , we assume the fading process  to be spatially uncorrelated, while the time correlation is dictated by the Doppler spread. In  \cite{WHinMatrix2} the DFDD technique is extended  to a matrix formulation with matrix prediction coefficients
 $\boldsymbol{P}=\left[\begin{matrix}
		\mathbf{P}_1,\cdots,\mathbf{P}_V
	\end{matrix}\right]$, 
	where $\mathbf{P}_i=p_i\mathbf{I}$ and $V$ is the prediction order. This matrix coefficient approach is effectively equivalent to using scalar coefficients, maintaining the simplicity and effectiveness of the prediction method.
	
\begin{figure}[htbp!]
		\centering
		\includegraphics[width=0.6\linewidth]{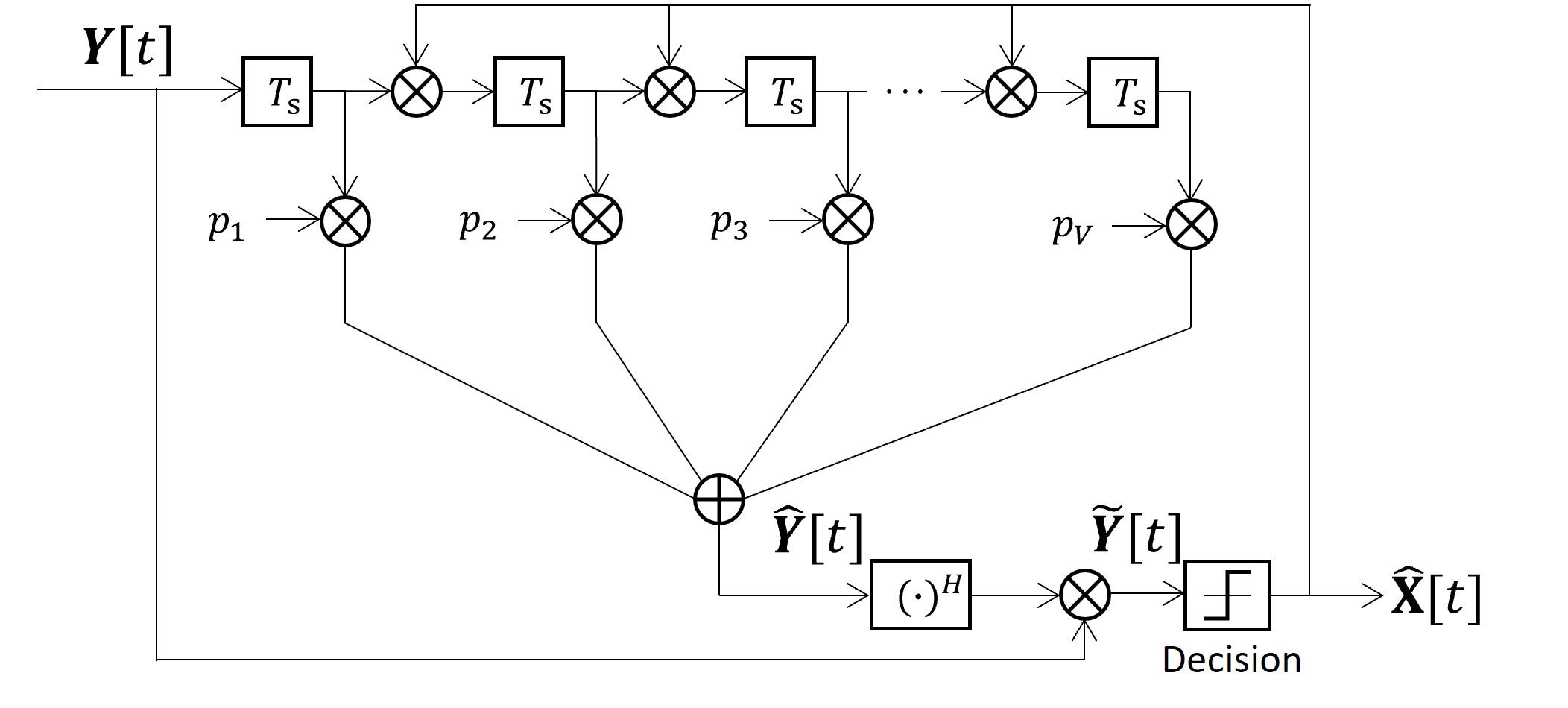}
		\caption{The structure of a DFDD receiver.}
		\label{DFDDreceiver}
	\end{figure}

	With  the DFDD technique, the receiver uses past decoded symbols  to assist in the detection of  the current symbol. Using (\ref{eq:diffencode}), then (\ref{eq:DFDDreceive}) can be expressed as \begin{equation}
		\boldsymbol{Y}[t]=\boldsymbol{\mathbf{H}}[t]\mathbf{V}[t-1]\mathbf{X}[t]+\mathbf{N}[t], \label{eq:Yt}
\end{equation}
Using  (\ref{eq:DFDDreceive}) at time instant $t-1$ we have
\begin {equation}
		\boldsymbol{Y}[t-1]=\boldsymbol{\mathbf{H}}[t-1]\mathbf{V}[t-1]+\mathbf{N}[t-1], \label{eq:Yt-1}
	\end{equation}
	Since $\boldsymbol{\mathbf{H}}[t]\approx\boldsymbol{\mathbf{H}}[t-1]$ due to the slow variation of the fading process, we have
	\begin{align}
		\boldsymbol{Y}[t]=\boldsymbol{Y}[t-1]\mathbf{X}[t]+\mathbf{N}[t],
	\end{align} if the effect of noise $\mathbf{N}[t-1]$ is neglected. 
	Then, $\boldsymbol{Y}[t-1]$ can be regarded as an estimate for $\boldsymbol{\mathbf{H}}[t]\mathbf{V}[t-1]$,  
	where the noise terms $\mathbf{N}[t-1]$  induce errors in these estimates. This effect can be reduced by using several past symbols  $\boldsymbol{Y}[t-v]$, and employing a smoothing procedure.
	The DFDD receiver relies on estimates for
	$\boldsymbol{\mathbf{H}}[t]\mathbf{V}[t-1]$ in (\ref{eq:Yt}) not only from $\boldsymbol{Y}[t-1]$ but also from the previous observed signals \cite{JiaweiThesis}.   Optimal linear prediction from previous $V$ signals $\boldsymbol{Y}[t-v],1\leq v\leq V$, for $\mathbf{H}[t]\mathbf{V}[t-1]$ can be  interpreted as resulting in a reference signal 
	$\hat{\boldsymbol{Y}}[t]$ for the detction of $ \boldsymbol{Y}[t]$, and it is formed as \cite{OnDFDD}
	\begin{align}
		\hat{\boldsymbol{Y}}[t] = \sum_{v=1}^{V}p_v\boldsymbol{Y}[t-v]\prod_{\mu=1}^{v-1}\hat{\mathbf{X}}[t-\mu], \label{eq:refsignal}
	\end{align}
	which it is illustrated in Fig. \ref{DFDDreceiver}.
	The product $\prod_{\mu=1}^{v-1}\hat{\mathbf{X}}[t-\mu]$ in Fig. \ref{DFDDreceiver} is formed from decision feedback symbols$\mathbf{\hat{X}}[t - (v - 1)]$ as
	\begin{align}
		\prod_{\mu=1}^{v-1}\hat{\mathbf{X}}[t-\mu]= \left\{\begin{matrix} {{\mathbf{\hat{X}}}[t - (v - 1)] \cdots {\mathbf{\hat{X}}}[t - 1],} & {v > 1} \\ {{{\mathbf{I}}_K},} & {v = 1}  \end{matrix} \right.
	\end{align}
	.
	
	Then, the decision variable for the DFDD receiver is formed as
	\begin{align}
		\tilde{\boldsymbol{Y}}[t] = \hat{\boldsymbol{Y}}^H[t]\boldsymbol{Y}[t],
\label{eq:DFDD_ref_symb}
	\end{align}
	and the decision rule is given by \cite{JiaweiThesis}
	\begin{align}
		\hat{\mathbf{X}}[t]=\arg\max\limits_{\mathbf{X}[t]\in\mathcal{X}} \Re \left[{\rm tr}\left(\mathbf{X}[t]\tilde{\boldsymbol{Y}}^H[t]\right)\right], \label{eq:simpleDFDD}
	\end{align}
	Using  (\ref{eq:DFDD_ref_symb}) in (\ref{eq:simpleDFDD}) and (\ref{eq:refsignal}) we have 
	\begin{align}
		\hat{\mathbf{X}}[t] =&\arg\max\limits_{\mathbf{X}[t]\in\mathcal{X}} \Re \left[{\rm tr}\left(\mathbf{X}[t]\boldsymbol{Y}^H[t]\hat{\boldsymbol{Y}}[t]\right)\right] \notag \\
		=& \arg\max\limits_{\mathbf{X}[t]\in\mathcal{X}} \Re \left[{\rm tr}\left(\mathbf{X}[t]\boldsymbol{Y}^{H}[t]\sum_{v=1}^{V}p_v\boldsymbol{Y}[t-v]\prod_{\mu=1}^{v-1}\hat{\mathbf{X}}[t-\mu]\right)\right],\label{eq:DFDD}
	\end{align}
	where $V$ is the prediction order, representing the total number of the previously detected symbols that are used to establish the reference $\hat{\boldsymbol{Y}}[t]$, and $p_v, v=1,\cdots, V$ are predictor coefficients determined by minimizing the mean-square error (MSE) between $\mathbf{H}[t]\mathbf{V}[t-1]$ and $\hat{\boldsymbol{Y}}[t]$,
	\begin{align}
		\sigma^2_{\rm mse} = E\big\{\big\lVert\mathbf{H}[t]\mathbf{V}[t-1]-\hat{\boldsymbol{Y}}[t]\big\rVert^2_F\big\}. \label{eq:mse1}
	\end{align}
	This criterion ensures that  $\hat{\boldsymbol{Y}}[t]$ is a good approximation to the current channel state $\mathbf{H}[t]\mathbf{V}[t-1]$ when using previous symbol decisions, thereby improving detection performance. In the derivation of  predictor coefficients  $p_v$ we assume previous decisions are correct
	as in  \cite{DFDDflatfading,linearpredictDFDD}. The derivation details are presented  in Appendix \ref{apx1}.	
	It is worth noting that for $V=1$, (\ref{eq:DFDD}) is the same as  conventional differential detection as long as the predictor coefficients are real.
	
	\section{Performance of DFDD-detected DRM }\label{section:3}
	To assess the performance of DFDD in DRM RIS  systems we perform extensive  Monte-Carlo simulations with BPSK and QPSK modulation. Different scenarios are considered, including various values for  $f_DT_s$, and $V$. 
In the simulations set up, the SNR is $\rho=\frac{1}{\sigma^2}$, where $\sigma^2$ is the variance of each component in the additive noise matrix.  For each SNR, we generate  $r\cdot10^{9}$ information bits, where $r$ is defined in (\ref{eq:r-bits}). We collect at least 50000-bit errors using Monte Carlo simulations on a space uncorrelated  time-varying Rayleigh fading channel based on Jakes' model, where $10^2$ blocks of information bits are randomly generated with a discrete uniform distribution, and then transmitted for each frame.
	
	\subsection{Results for $2\leq K\leq4$ and BPSK}\label{section:DFDDM2}
	We start with  the error performance of the DFDD receiver for BPSK. To evaluate how different parameters affect the system, in our simulations we vary $f_DT_s$ from $0.01$ to $0.03$ and change the prediction order $V$ from $1$ to $3$. Using the system illustrated in Fig. \ref{DFDDsystem}, we generate the fading processes using Jakes' model and then proceed with the modulation and perform  decoding using the  DFDD technique.

	\begin{figure}[htbp]
		{
			\begin{minipage}[t]{0.5\linewidth}
				\centering         
				\includegraphics[width=\linewidth]{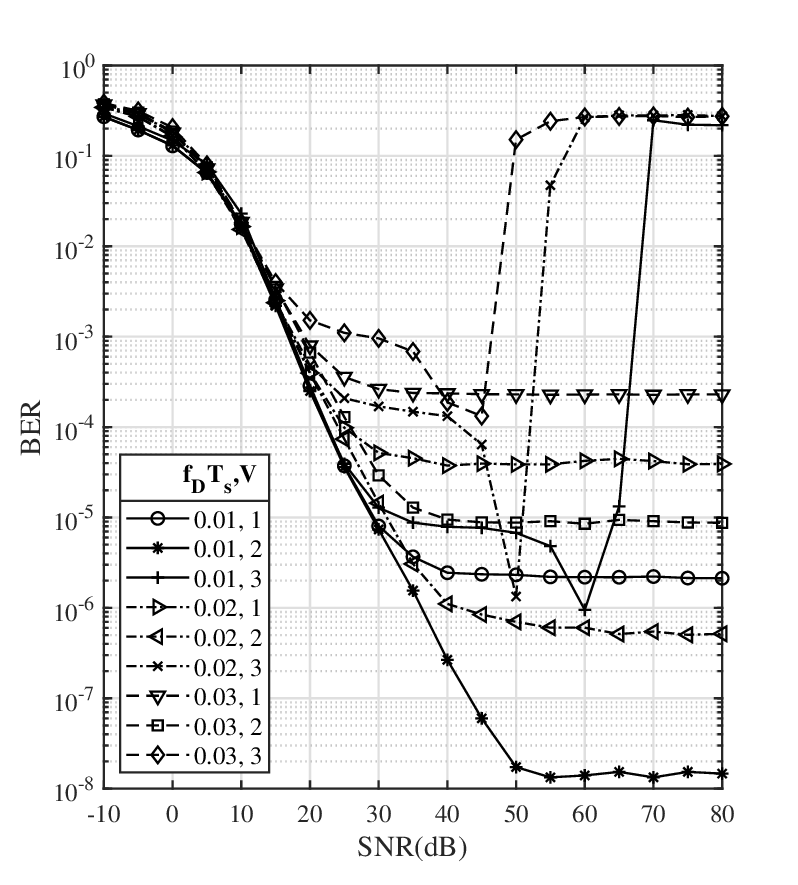}
				\caption{BER performance of the DFDD-detected DRM system with BPSK when $K=2$, $f_DT_s=0.01,0.02,0.03$, and $V=1,2,3$.}
				\label{BK2}
			\end{minipage}
		}
		{
			\begin{minipage}[t]{0.5\linewidth}
				\centering      
				\includegraphics[width=\textwidth]{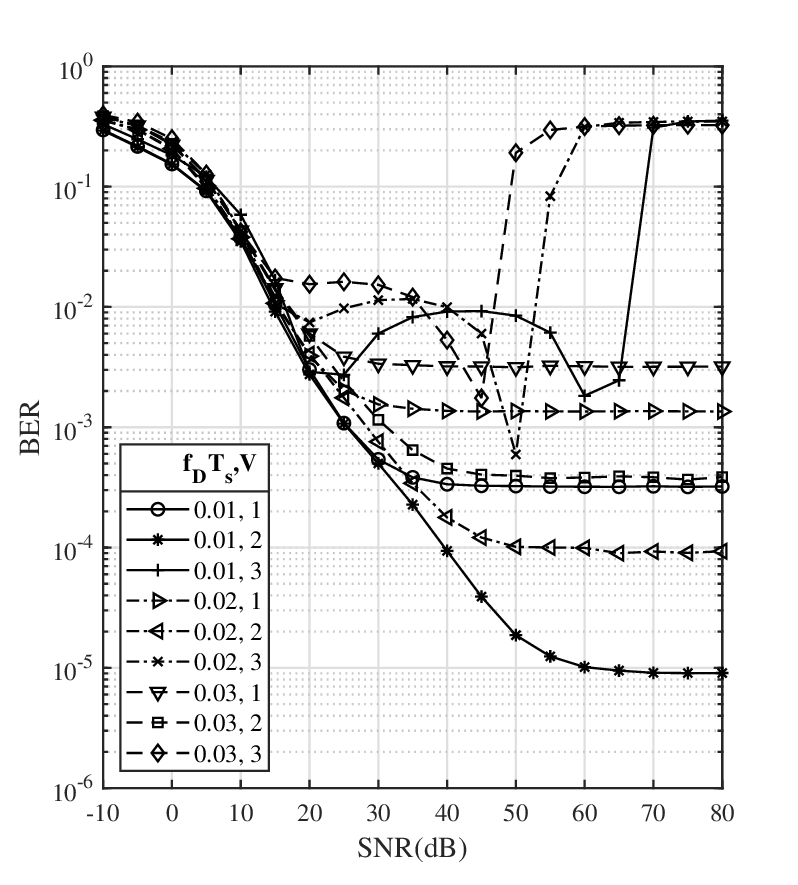}
				\caption{BER performance of the DFDD-detected DRM system with BPSK when $K=3$, $f_DT_s=0.01,0.02,0.03$, and $V=1,2,3$.}
				\label{BK3}
			\end{minipage}
		} 
	\end{figure}
	
	In Fig. \ref{BK2} iwe present the system performance for $K=2$. As the normalized Doppler frequency $f_DT_s$ increases from 0.01 to 0.03, there is a noticeable decline in performance across all prediction orders. At a normalized Doppler frequency of 0.01, the system achieves its lowest BER of $2.5\cdot10^{-6}$ at an $E_b/N_0$ of 40 dB with $V=1$. However, when $f_DT_s$ rises to 0.02, the BER encounters an error floor of approximately $4\cdot10^{-5}$ at 35 dB SNR, and at $f_DT_s=0.03$, the error floor further increases to around $2.4\cdot10^{-4}$ at 30 dB SNR. This performance degradation persists for all prediction orders as the normalized Doppler frequency increases. The reason for this is that as $f_DT_s$ grows, the channel varies more rapidly, and hence does nor remains the same over past transmitted symbols.
 The occurrence of an error floor is often attributed to the high error probability associated with the first few decoded symbols. These symbols are decoded by CDD rather than DFDD \cite{JunqianThesis} due to the insufficient availability of previously decoded symbols for the scheme.
 However, $V=2$ provides the lowest error floors. for all values of normalized Doppler, indicating that DFDD has the property of lowering the error floor.
 	
	In Fig. \ref{BK2}, for a specific $f_D T_s$, it can be seen that at low SNR, the BER for $V=2$ is slightly  better than for $V=1$. As the SNR rises, the error performance of $V=2$ becomes noticeably better than that of $V=1$. For example, in Fig. \ref{BK2}, when the normalized Doppler frequency is 0.02, the gap between $V=2$ and $V=1$ at a BER of $10^{-4}$ is about 1 dB. As the SNR continues to increase, the BER for $V=1$ exhibits  an error floor around $4 \cdot 10^{-5}$ starting at the SNR of 35 dB. In contrast, the BER for $V=2$ keeps improving, dropping to $ 10^{-6}$ at $E_b/N_0 = 40$ dB.
	Across all normalized Doppler frequencies, it is evident that $V=2$ consistently outperforms $V=1$. The higher prediction order $V$ provides more accurate  reference detection symbol (\ref{eq:refsignal})  improving  current symbol detection, and resulting in better performance. 

	 As illustrated in Fig. \ref{BK2}, a larger prediction order may degrade performance. For example, when $f_D T_s = 0.01$ and $V = 3$, the BER reaches a minimum value of $10^{-6}$ at an $E_b/N_0$ of 60 dB. However, before this point, the performance is consistently worse compared to that of $V = 1$ and same $f_D T_s$ . Then, the error rate rapidly increases in the higher SNR range. For $V = 3$ and $f_D T_s = 0.02$, the BER attains its lowest point at $1.5 \cdot 10^{-6}$ and then sharply increases at  $E_b/N_0$ larger than 50 dB. This phenomenon  occurs even earlier for $f_D T_s = 0.03$, with the BER reaching $1.4 \cdot 10^{-4}$ at the SNR of 45 dB. Such  trends of BER behaviour can be observed in all scenarios for $V = 3$. 	
	With a higher prediction order, the DFDD receiver employs a larger number of previously detected symbols increasing the risk of error propagation. If a previous decision is incorrect, it persist a laonger time in the receiver and hence it has a more significant impact in demodulation, possibly resulting in increasing BEDR. Such a phenomenon for scalar DFDD receivers has been considered in detail in \cite{Junqian}. Therefore for any values of $V$, performance significantly deteriorates under high Doppler frequencies, making it essential to select an optimal prediction order $V$ for best performance.
	
\begin{figure}[htbp]
		{
			\begin{minipage}[t]{0.5\linewidth}
				\centering         
				\includegraphics[width=\linewidth]{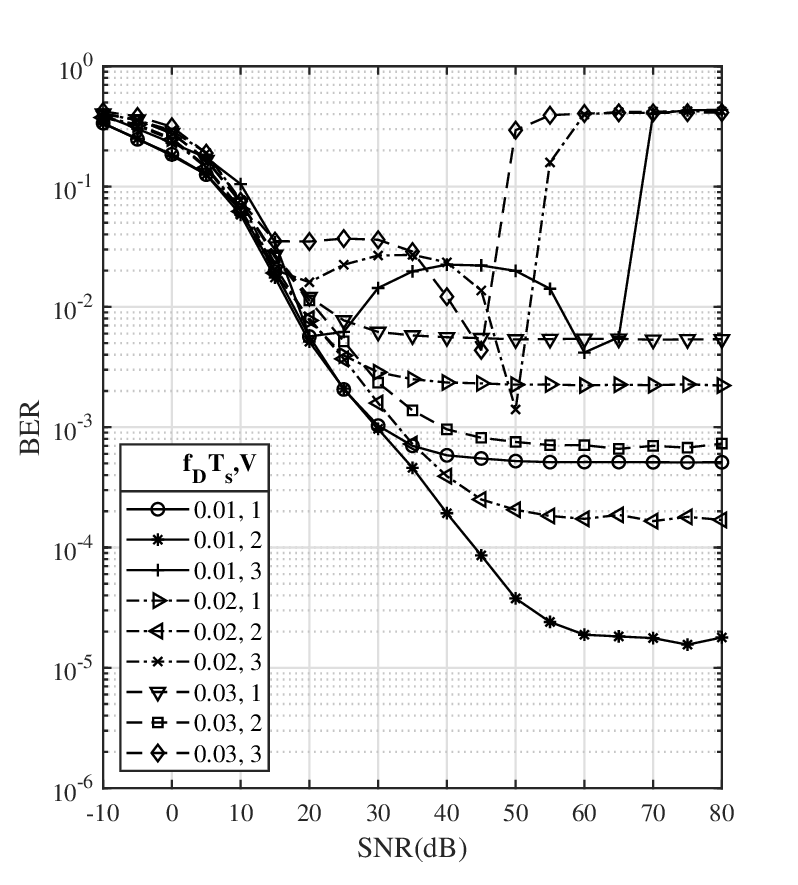}
				\caption{BER performance of the DFDD-detected DRM system with BPSK when $K=4$, $f_DT_s=0.01,0.02,0.03$, and $V=1,2,3$.}
				\label{BK4}
			\end{minipage}
		}
		{
			\begin{minipage}[t]{0.5\linewidth}
				\centering      
				\includegraphics[width=\textwidth]{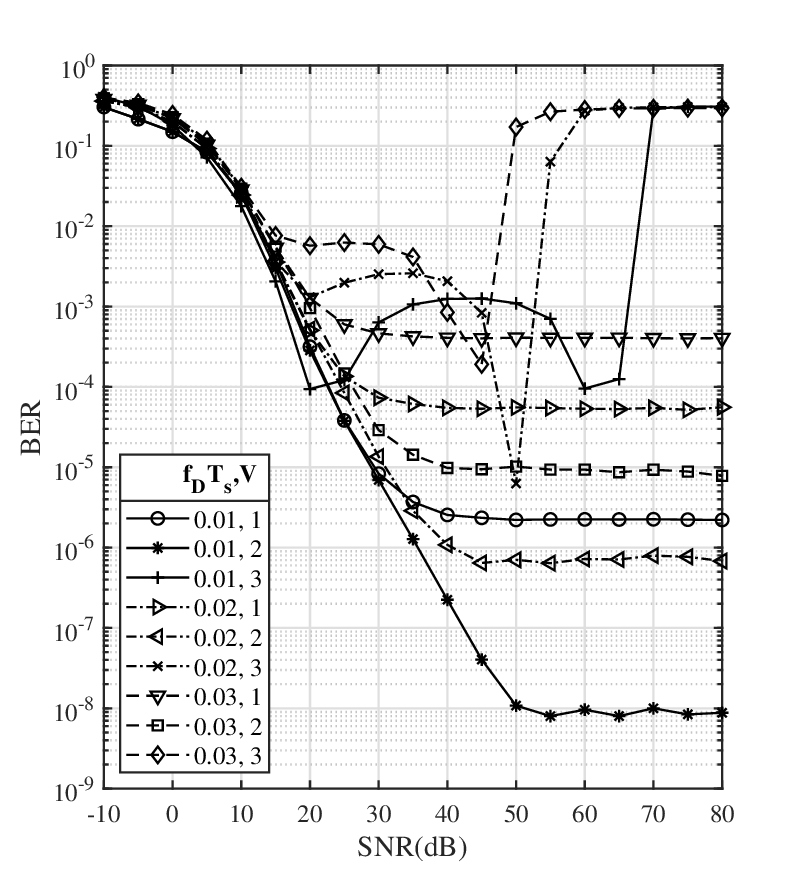}
				\caption{BER performance of the DFDD-detected DRM system with QPSK when $K=2$, $f_DT_s=0.01,0.02,0.03$, and $V=1,2,3$.}
				\label{QK2}
			\end{minipage}
		} 
	\end{figure}

	Following the performance analysis  of the DFDD receiver for $K=2$, we now consider scenarios where $K\geq2$ with BPSK Since increasing $K$ leads to a greater complexity, our evaluations primarily involve $K=3$ and $K=4$, of which the BER performance  is illustrated in Fig. \ref{BK3} and Fig. \ref{BK4}, respectively.
	These figures generally indicate similar error behaviour as observed for $K=2$. For a given prediction order $V$, the system performance worsens as the normalized Doppler frequency increases. For example, when $K=4$ and $V=1$, an error floor of $5\cdot10^{-4}$ is reached at $E_b/N_0 > 40$ dB for $f_DT_s = 0.01$, while for $f_DT_s = 0.03$, the minimum BER degrades to $5.5\cdot10^{-3}$ at $E_b/N_0 > 30$ dB. Across all Doppler frequencies, using $V=2$ consistently leads to better performance compared to $V=1$ throughout the SNR range, while larger prediction orders such as $V=3$ result in performance degradation. For instance, for $K=3$ and $f_DT_s=0.01$, $V=2$ yields a s better performance  than $V=1$ by about 1 dB at a BER of  $10^{-3}$ . With $V=2$, the BER encounters an error floor of $10^{-5}$ at an SNR of  60 dB , which is significantly better than the $3.2\cdot10^{-4}$ error floor for $V=1$ at $E_b/N_0\geq30$ dB. In contrast, for $f_DT_s = 0.03$ and $V=3$ in Fig. \ref{BK3}, the BER initially drops to $1.8\cdot10^{-3}$ at 45 dB but quickly increases again. Throughout the entire SNR range, $V=3$ demonstrates worse performance than $V=2$, with a pronounced difference of 20 dB at a BER of $10^{-2}$.
	
	When comparing Fig. \ref{BK2}, Fig. \ref{BK3}, and Fig. \ref{BK4}, with $V$ and $f_DT_s$ constant, it is evident that the error performance deteriorates as $K$ increases,. For example, when $f_DT_s = 0.03$ and $V=2$, the error floor is approximately $9\cdot10^{-6}$ for $K=2$, $4\cdot10^{-4}$ for $K=3$, and $7\cdot10^{-4}$ for $K=4$. This increase in BER floor as $K$ increases emphasizes the more significant impact of larger $K$ on system performance.

	\subsection{Results for DFDD detection with $2\leq K\leq4$ and QPSK}
	We consider now  the performance of DFDD receivers in DRM RIS systems with QPSK for  $2\leq K\leq4$, and compare it to the BPSK case. The normalized Doppler frequency $f_DT_s$ ranges from $0.01$ to $0.03$, and the prediction order $V$ assumes the values $1$, $2$ and $3$.
	
\begin{figure}[htbp]
		{
			\begin{minipage}[t]{0.5\linewidth}
				\centering         
				\includegraphics[width=\linewidth]{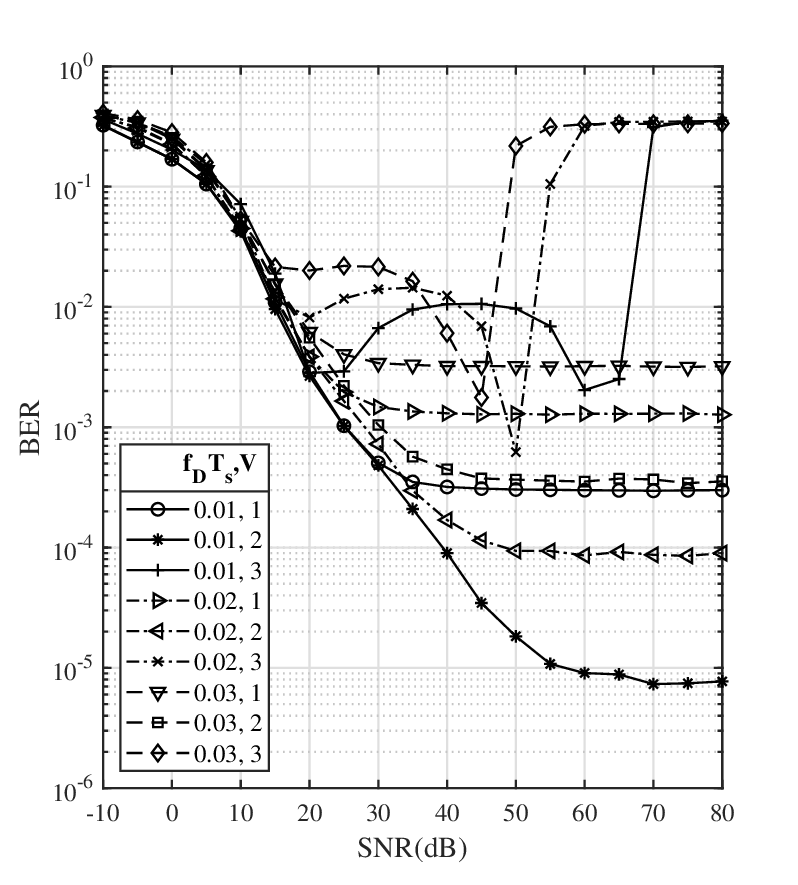}
				\caption{BER performance of the DFDD-detected DRM system with QPSK when $K=3$, $f_DT_s=0.01,0.02,0.03$, and $V=1,2,3$.}
				\label{QK3}
			\end{minipage}
		}
		{
			\begin{minipage}[t]{0.5\linewidth}
				\centering      
				\includegraphics[width=\textwidth]{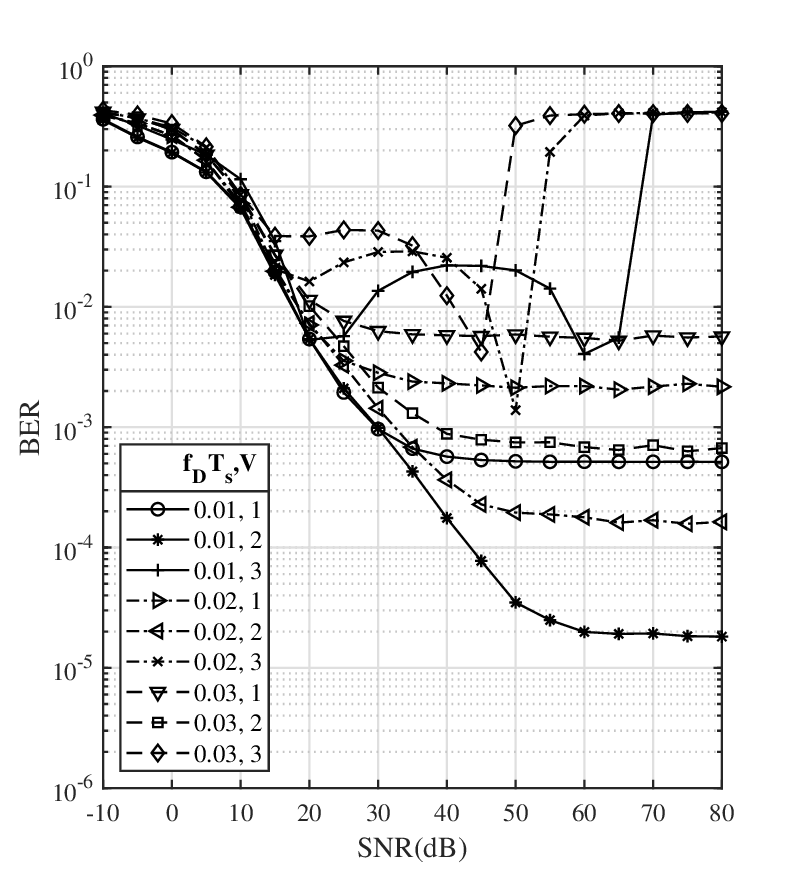}
				\caption{BER performance of the DFDD-detected DRM system with QPSK when $K=4$, $f_DT_s=0.01,0.02,0.03$, and $V=1,2,3$.}
				\label{QK4}
			\end{minipage}
		} 
	\end{figure}

	 As depicted in Figs. \ref{QK2}, \ref{QK3}, and \ref{QK4}, the general behavior with QPSK is similar to that of the BPSK case discussed earlier. A key observation is the rising BER error floor as the Doppler frequency increases, suggesting that QPSK is more sensitive to fast channel variations. For example, in Fig. \ref{QK2}, when $K=2$ and $V=1$, for $f_D T_s=0.01$. the BER reaches an error floor of approximately $2.2\cdot10^{-6}$  at  $E_b/N_0\geq 40$ dB. When the Doppler frequency increases to 0.02, the error floor rises to about $5.5\cdot10^{-5}$ at SNR values above 30 dB.
Additionally, with a prediction order of $V=2$, the BER performance closely aligns with that of $V=1$ at lower SNR values. However, as the SNR increases, $V=2$ presents significantly better performance than $V=1$. As the SNR increases, $V=1$ reaches an error floor around $5\cdot10^{-5}$ at $E_b/N_0\geq40$ dB, while the BER for $V=2$ continues to drop, approaching $6\cdot10^{-7}$ at the SNR of 45 dB.
	
	The phenomenon of increased BER at high SNR, observed for BPSK  appears also for QPSK, especially when the prediction order is 3. As shown in Fig. \ref{QK4}, for $f_D T_s=0.02$ and $V=3$, the BER reaches a minimum of $1.5\cdot10^{-3}$ at an SNR of 50 dB  before rising sharply. Except for this minimum, $V=3$ consistently performs worse compared to $V=1$. Moreover, as the Doppler frequency increases, the SNR, at which the BER starts to increase rapidly, decreases. For example, with $f_DT_s=0.01$ and $V=3$, the BER attains a minimum of $4\cdot10^{-3}$ at $E_b/N_0=60$ dB, whereas for $f_DT_s=0.03$, the minimum BER of $4\cdot10^{-3}$ occurs at 45 dB SNR. However, it is seen that also in this case the performance is best when $V=2$, providing significant advantages over $V=1$.
	
	We next compare the QPSK performance with the BPSK results presented previously. By comparing Fig. \ref{QK2} with Fig. \ref{BK2}, it is clear that the error floors for QPSK are approximately the same or  only slightly higher than those for BPSK.  When the normalized Doppler frequency is 0.03 and the prediction order is 1, BPSK reaches an error floor at a BER of about $2.3\cdot10^{-4}$, while QPSK encounters a slightly higher floor at $4\cdot10^{-4}$. However, in most cases, the performance of BPSK and QPSK is quite close, particularly in the high SNR range. This similarity in behavior can also be seen in the DRM schemes over slower fading channels. For example, in Figs. \ref{BK4} and \ref{QK4}, for a normalized Doppler frequency of 0.01 and a prediction order of $V=2$, both constellations reach an error floor around $1.8\cdot10^{-5}$ at $E_b/N_0\geq60$ dB.
	
	To examine the performance when $V=3$ and $f_DT_s=0.02$, Fig. \ref{QK2} shows that with QPSK the lowest BER point, where the BER starts increasing, occurs at around $6\cdot10^{-6}$ at $E_b/N_0=50$ dB. In comparison, Fig. \ref{BK2} demonstrates that BPSK achieves a lower minimum BER of approximately $1.5\cdot10^{-6}$ at the same $E_b/N_0$ value, indicating a slight advantage for BPSK over QPSK in this context.

	\section{Comparison with Conventional Differential Detection (CDD) of  DRM and Complexity Analysis }\label{section:4}
	In this section, we compare the performance of the proposed DFDD-detected DRM scheme with the original DRM system as well as  with  Differential Space-Time Modulation (DSTM) coded DRM scheme from  \cite{qiu2026space} using CDD.  It was shown in \cite{qiu2026space}, that over fading time varying channels, DRM with DSTM can also reduce error floors. Hence from this posint of viewe it is interesting to compare with uncoded DRM with DFDD which was shown to have same proprty.Then, the  complexity of these systems is analyzed.
	
	\subsection{Performance Comparison}	
	
	 Results for uncoded DRM and DRM-DSTM systems from \cite{qiu2026space}, which both use CDD,  are compared with DRM with DFDD for $K=2,3,4$ over Jake's fading channels. The notation for DSTM coded DRM follows the same format as in\cite{qiu2026space}. IIn this section we select the parameters that provide the best performance at comparable levels of complexity as found in our work. For instance, choosing $V=2$ for DFDD provides the best performance, particularly at high SNR.

	\begin{figure}[htbp!]
		{
			\begin{minipage}[t]{0.5\linewidth}
				\centering
				\includegraphics[width=\linewidth]{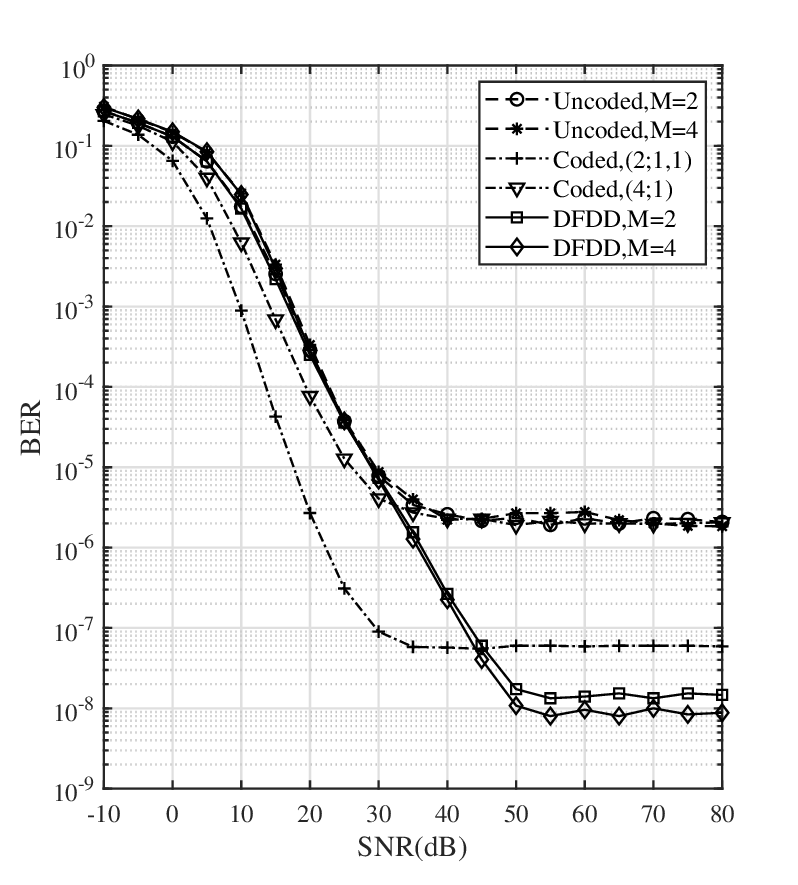}
				\caption{BER performance of  DRM,, DRM-DSTM with CDD and DRM with  DFDD for  $K=2$. and $f_DT_s=0.01$}
				\label{K2cc}
			\end{minipage}
		}
		{
			\begin{minipage}[t]{0.5\linewidth}
				\centering
				\includegraphics[width=\linewidth]{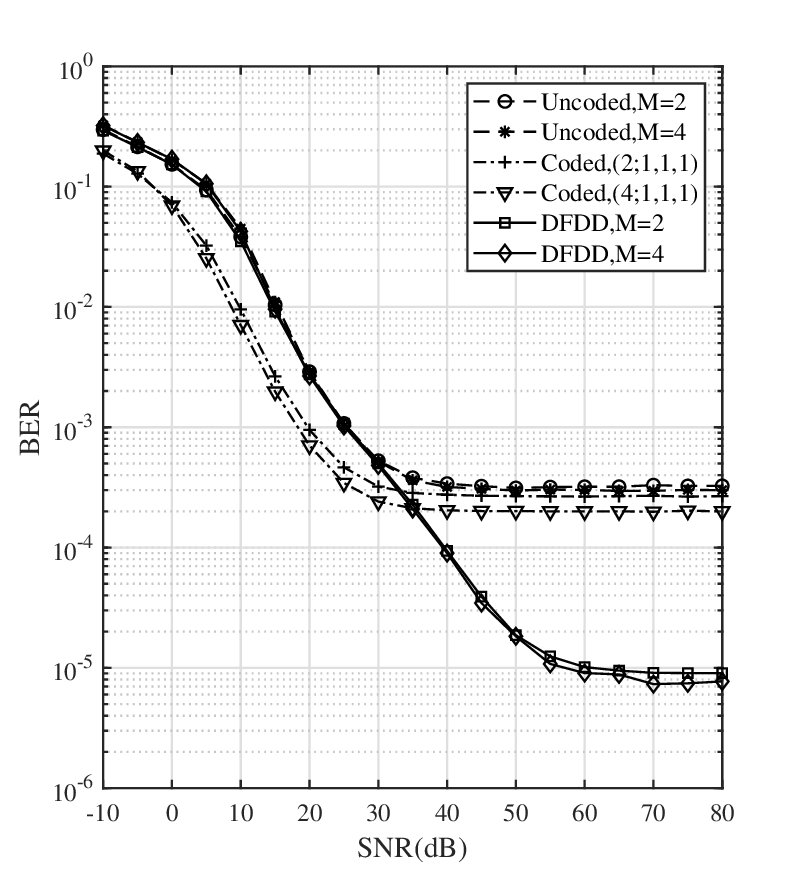}
				\caption{BER performance of  DRM,, DRM-DSTM with CDD and DRM with  DFDD for  $K=3$. and $f_DT_s=0.01$}
				\label{K3cc}
			\end{minipage}
		} 
	\end{figure}

\begin{figure}[htbp!]
		\centering
		\includegraphics[width=0.5\linewidth]{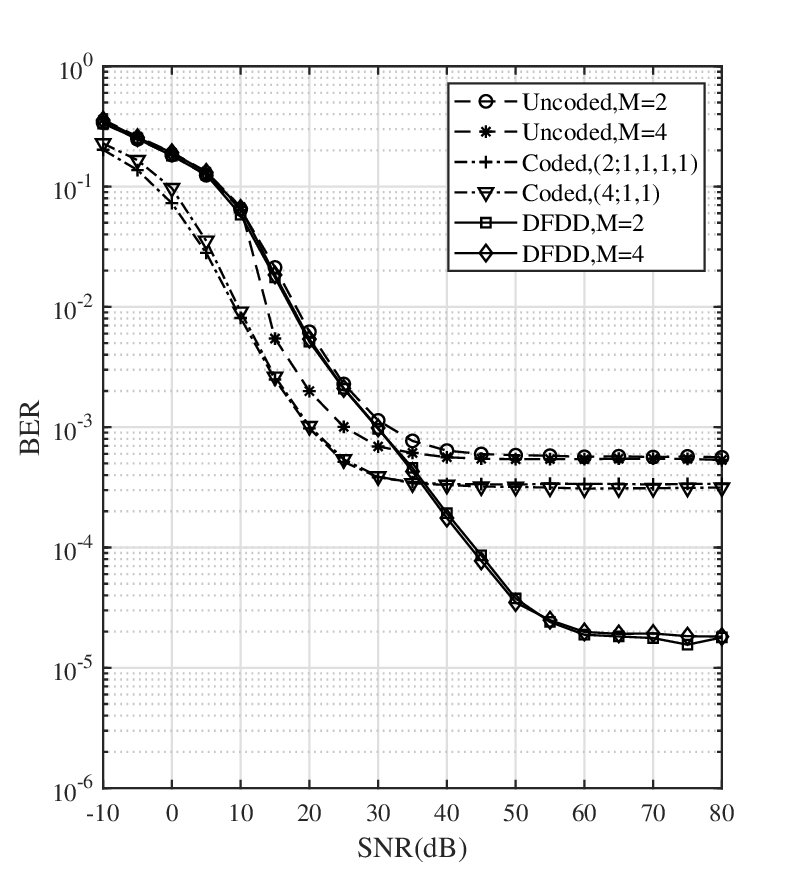}
		\caption{BER performance of  DRM,, DRM-DSTM with CDD and DRM with  DFDD for  $K=4$. and $f_DT_s=0.01$}
			\label{K4cc}
		\end{figure}

	In Fig. \ref{K2cc}, \ref{K3cc}, and \ref{K4cc} we illustrate the performance of DRM, DRM-DSTM with CDD, and DRM with DFDD,  over time varying channels with  $f_DT_s=0.01$.
	We note a clear  performance difference between these three schemes, however in all these cases we see a marked error floor effect. From \cite{qiu2026space}  we have that over quasi-static fading channels, employing suitable cyclic or dicyclic codes in DRM-DSTM  significantly enhances performance relative to uncoded DRM . This benefit of DSTM is also evident under time-varying conditions. For example, in Fig. \ref{K3cc}, with $K=3$ and $f_DT_s=0.01$, the $(4;1,1,1)$-coded system improves performance by about 6 dB over the uncoded DRM with QPSK at a BER of $10^{-3}$. At $E_b/N_0\geq30$ dB, the $(2;1,1,1)$-coded and $(4;1,1,1)$-coded schemes are impaired by error floors of $2\cdot10^{-4}$ and $3\cdot10^{-4}$, respectively, whereas the original DRM scheme achieves a minimum BER of $4.2\cdot10^{-4}$. This performance gap widens with increasing $K$, as shown by the approximately 11 dB difference between the $(2;1,1,1,1)$ coded system and the uncoded BPSK scheme at a BER of $10^{-3}$ in Fig. \ref{K4cc}, compared to a 7 dB gap between the $(2;1,1)$ cyclic-coded scheme and the uncoded scheme in Fig. \ref{K2cc}.

Furthermore, for $K=4$, the performance of the cyclic and dicyclic coded systems is nearly identical. For instance, in Fig. \ref{K4cc} with $f_DT_s=0.01$, the performance difference between cyclic $(2,1,1,1,1)$ and dicyclic $(4,1,1)$ codes is negligible, with both curves overlapping almost entirely across the SNR range. In contrast, for $K=2$, Fig. \ref{K2cc} reveals a substantial performance discrepancy between two types of  the group codes. Specifically, at a BER of $10^{-5}$, the $(2;1,1)$ cyclic-coded scheme  outperforms the $(4;1)$ dicyclic-coded scheme by about 8 dB. The $(2;1,1)$-coded scheme reaches an error floor of $5.9\cdot10^{-8}$ at $E_b/N_0\geq35$ dB, whereas the $(4;1)$-coded scheme shows an error floor of $2.5\cdot10^{-6}$ for SNR above 35 dB.
	
	Next we  compare the performance of the DFDD system with the original DRM with DRM- DSTM schemes. As observed, increasing Doppler frequency generally degrades performance across all SNR ranges. Therefore, we focus on $f_DT_s=0.01$ for clarity in the figures. It is evident that while DFDD-based DRM schemes initially lag behind DSTM-coded schemes, they eventually surpass them at high SNR levels. Compared to the original DRM schemes, DFDD schemes perform similarly at low SNR but performs better  as SNR increases.
	In Fig. \ref{K2cc}, at a BER of $10^{-4}$, the DFDD for $f_DT_s=0.01$ lags approximately  by 3.5 dB behind the $(2;1,1)$-coded scheme and shows almost no difference from the uncoded DRM. As SNR rises, this gap grows; for example, at a BER of $10^{-6}$, the difference between DFDD and DSTM systems increases to about 11.5 dB. The poorer performance of DFDD at low SNR is attributed  to more errors being propagated  in the decision feedback process, as discussed in section \ref{section:DFDDM2}.
	However, as SNR increases, the advantages of DFDD systems become more apparent.  The CDD-detected DRM scheme reaches an error floor of approximately $2.5\cdot10^{-6}$ at $E_b/N_0\geq40$ dB, and it falls significantly behind  DFDD-detected schemes for both BPSK and QPSK. The BER of the DFDD scheme continues to improve until it reaches an error floor around $10^{-8}$. Similarly to DRM  with BPSK and QPSK, the DFDD scheme exhibits nearly identical BER performance for both signal constellations at high SNR. In Fig. \ref{K2cc}, for instance, the $(2;1,1)$-coded scheme encounters an error floor of $5.7\cdot10^{-8}$ at $E_b/N_0\geq35$ dB, while the BER of DFDD with  uncoded BPSK keeps decreasing, eventually reaching a lower error floor of $1.5\cdot10^{-8}$ at 50 dB SNR. The DFDD scheme with uncoded BPSK outperforms the $(2;1,1)$-coded scheme at an SNR of around 45 dB, with a similar crossover occurring earlier for QPSK at an SNR of 32 dB.
	
	These trends are also observed in other figures with different $K$ values. In Fig. \ref{K3cc}, both BPSK and QPSK show that DFDD starts to outperform DRM schemes at $E_b/N_0\geq30$ dB, although both systems perform similarly at low SNR. The DFDD scheme, for example, is impaired by an error floor of approximately $10^{-5}$, which is significantly lower than the DRM error floor of $4.2\cdot10^{-4}$. While the DSTM-coded scheme initially performs better, it is eventually surpassed by the DFDD system at medium to high SNR and reaches a higher error floor compared to DFDD. For instance, the $(4;1,1,1)$-coded scheme with QPSK shows a 6 dB advantage at a BER of $10^{-3}$ at low SNR but reaches an error floor of $2\cdot10^{-4}$ at $E_b/N_0 \geq 35$ dB, while the BER continues to improve for DFDD, reaching an an error floor of  $8\cdot10^{-6}$ beyond 55 dB.
	
	Despite the overall degradation in error rate performance with increasing $K$, the trends we discussed before remain the same. For $K=4$ in Fig. \ref{K4cc}, a clear performance gap emerges between DFDD and DRM schemes after $E_b/N_0 = 30$ dB, with the two schemes reaching error floors of approximately $2\cdot10^{-5}$ and $5.7\cdot10^{-4}$, respectively. The $(2;1,1,1,1)$ cyclic-coded scheme outperforms  DFDD with QPSK by about 10 dB at a BER of $10^{-3}$. However, an error floor of $3.7\cdot10^{-4}$ at an SNR of 30 dB SNR impaires the coded scheme, while the DFDD scheme eventually achieves an error floor of $2\cdot10^{-5}$ at $E_b/N_0 \geq 60$ dB.

	\subsection{Complexity Analysis}
	This section examines the encoding and decoding complexity of the DFDD system, comparing it with the DSTM-coded system and the uncoded DRM scheme both with CDD receivers.
	
	First, the common processes of generating permutation matrices, mapping $M$-PSK symbols, and obtaining information-carrying matrices, as used in the DRM scheme, are omitted. The differential encoding described in (\ref{eq:diffencode}) has a time complexity of $O(K^3)$. The number of multiplications involved during modulation in (\ref{eq:DFDDreceive}) is $K^2\cdot N_r\cdot K=N_rK^3$, making the overall  complexity to be $O(K^3)$. A summary of the time complexity for each encoding step is provided in Table \ref{tb:DFDDencodecomplexity}.
	
	\begin{table}[!htb]
		\centering
		\caption{Time complexity of each step in the encoding process of DFDD-detected DRM scheme}
		\begin{tabular}{ccccc}
			\hline
			& Multiplication & Addition & Copy & Time Complexity\\
			\hline
			Generate permutation matrix & - & - & - & $O(\binom{K!}{2^{r_1}})$\\
			M-PSK symbol mapping & - & - & - &  $O(K\log_2M)$\\
			Generate information-carrying matrix & $K^3$ & - & - & $O(K^3)$ \\
			Differential Encoding & $K^3$ & - & - & $O(K^3)$\\
			Generate channel matrix & $K^2N^2N_r+K^2NN_r$ & $N_rK$ & - &   $O(K^2)$\\
			Modulation & $K^2N_r$ & $N_rK$ & - & $O(K^2)$\\
			
			\hline
		\end{tabular}
		\label{tb:DFDDencodecomplexity}
	\end{table}
	
	During the detection, following (\ref{eq:DFDD}), for each candidate $\mathbf{X}[t]\in\mathcal{X}$, $\sum_{v=1}^{V}((v-1)K^3+K^2N_r)+K^2N_r+K^3$ multiplications are required for matrices multiplication, resulting in the time complexity being $O(VK^3)$. By iterating over $\lvert\mathcal{X}\rvert=2^r$ candidates, the time complexity of the whole demodulation process is $O(2^rVK^3)$. The computing of trace includes $K$ additions and the time complexity is $O(K)$.
	
	\begin{table}[!htb]
		\centering
		\caption{Detector complexity comparison}
		\begin{tabular}{ccc}
			\hline
			Detector & Numbers of Multiplications & Time Complexity\\
			\hline
			Uncoded DRM with CDD & $K^2N_r+K^3$ & $O(2^rK^3)$\\
			DSTM-coded DRM with CDD & $K^2N_r+K^3$ & $O(2^{r^{\prime}}K^3)$\\
			Uncoded DRM with DFDD & $\sum_{v=1}^{V}((v-1)K^3+K^2N_r)+K^2N_r+K^3$ & $O(2^rVK^3)$ \\
			\hline
		\end{tabular}
		\label{tb:complexity compare}
	\end{table}
	
	In conclusion, the main difference in time complexity of the DFDD-detected system with respect to the uncoded DRM scheme and DSTM-coded system with CDD is the detection process. Table \ref{tb:complexity compare} presents the numbers of multiplications required for each system during detection, which indicate the difference between the three shemes. It is seen that the DFDD complexity depends linearly on the predictor order $V$. The DSTM coded DRM has the largest time complexity due to the term $2^{r^l}$.

	\section{Conclusions}\label{section:Conclusion}

	In this paper, we consider the use of a DFDD technique, not requiring channel state information for detction  in DRM RIS systems. The use of CDD,, which is more common in such systems, results in performance loss and error floors over time-varying fading channels, while DFDD based receivers address these shortcomings at the expense of a modest increased complexity. In this paper we present a  mathematical derivation of the  DFDD receiver for the DRM RIS scheme with suitable  simplifications. Monte-Carlo simulation results over Jakes' fading channels are used to assess performance. We show that the best performance for DFDD receivers in our RIS systems is obtained when the DFDD prediction order is 2.
 When comparing DRM with DFDD receivers to  DRM emplying CDD receivers over time-varying fading channels, DFDD shows comparable performance at low SNR. However, in the medium SNR range, DFDD continues to improve its error rate, whereas the DRM scheme is impaired by a relatively high error floor. The clear advantage of DFDD receivers for  DRM over CDD receivers is the significantly lower error floor over time varying fading channels. Although  DRM-DSTM coded schemes with CDD receivers also lower the error floor when compared to unocded DRM with CDD, this work shows that uncoder DRM with DFDD still has the lowest error floor among the three schemes we considered. The chief merits of the DFDD technique for DRM RIS systems is its significantly lower error floors over fading time varying channelsm even when the prediction order is 2, as used in our paper.
 Directions for further research on this subject could be the use of DFDD  techniques for coded DRM-DSTM systems. The use of DFDD  is a challenging research direction even for scalar channels, however it is envisaged that it could generate useful results. Another future research direction could be to develop alternate designs for the prediction coefficients of DFDD receivers following the approach from \cite{zhang2020regularized}, which has to be adapted to DRM RIS systems. This can allow the increase of the prediction order beyond 2,  reducing the decision feedback error propagation effects, and ultimately improving performance.

	\appendices
	\section{Derivation of Predictor Coefficients}\label{apx1}
	In this appendix, we present the derivation of the predictor coefficients $p_v$ from (\ref{eq:refsignal}). Minimizing (\ref{eq:mse1}) is equivalent to minimizing
	\begin{align}
		\sigma^2_{\rm mse} &= E\big\{\big\lVert\mathbf{H}[t]\mathbf{V}[t-1]-\hat{\boldsymbol{Y}}[t]\big\rVert^2_F\big\} 
		=E\big\{\big\lVert\boldsymbol{Y}[t]\mathbf{X}^{-1}[t]-\mathbf{N}[t]\mathbf{X}^{-1}[t]-\hat{\boldsymbol{Y}}[t]\big\rVert^2_F\big\} \notag\\
		&=E\big\{\big\lVert\boldsymbol{Y}[t]-\mathbf{N}[t]-\hat{\boldsymbol{Y}}[t]\mathbf{X}[t]\big\rVert^2_F\big\} \label{eq:mse2}
	\end{align}
	since (\ref{eq:Yt}) can be further expressed as
	\begin{align}
		\boldsymbol{Y}[t]-\mathbf{N}[t]&=\boldsymbol{\mathcal{H}}[t]\mathbf{V}[t-1]\mathbf{X}[t], \\
		(\boldsymbol{Y}[t]-\mathbf{N}[t])\mathbf{X}^{-1}[t]&=\boldsymbol{\mathcal{H}}[t]\mathbf{V}[t-1]. \label{eq:Xinvertible}
	\end{align}
	In (\ref{eq:Xinvertible}), $\mathbf{X}[t]$ is invertible because it is a product of a permutation matrix and a diagonal matrix with diagonal entries taking values over $M$-PSK constellations. It is known that a permutation matrix is also orthogonal, which is necessarily invertible with $\mathbf{Z}^{-1}[t]=\mathbf{Z}^{T}[t]$. A diagonal matrix, is invertible as long as its diagonal elements are all nonzero. Its inverse is another diagonal matrix with reciprocals of the original diagonal entries. As a product of two invertible matrices, $\mathbf{X}[t]$ is invertible as well.
\enlargethispage*{10mm}
	Without decision feedback errors, $\hat{\mathbf{X}}[t-v]=\mathbf{X}[t-v],1\leq v \leq V-1$, yielding
	\begin{align}
		\prod_{\mu=1}^{v-1}\hat{\mathbf{X}}[t-\mu]\mathbf{X}[t]=\prod_{\mu=1}^{v-1}\mathbf{X}[t-\mu]\mathbf{X}[t-0]=\prod_{\mu=0}^{v-1}\mathbf{X}[t-\mu].
	\end{align} In addition, we have
	\begin{align}
		\mathbf{V}[t]=\mathbf{V}[t-v]\prod_{\mu=0}^{v-1}\mathbf{X}[t-\mu]\notag\\
		\mathbf{V}^{-1}[t-v]\mathbf{V}[t]=\prod_{\mu=0}^{v-1}\mathbf{X}[t-\mu].
	\end{align}
	Therefore, (\ref{eq:mse2}) can be further expressed as
	\begin{align}
		\sigma^2_{\rm mse} &= E\big\{\big\lVert\boldsymbol{Y}[t]-\mathbf{N}[t]-\hat{\boldsymbol{Y}}[t]\mathbf{X}[t]\big\rVert^2_F\big\} \notag \\
		&=E\bigg\{\bigg\lVert \mathbf{H}[t]\mathbf{V}[t]+\mathbf{N}[t]-\mathbf{N}[t]-\bigg(\sum_{v=1}^{V}p_v\boldsymbol{Y}[t-v]\prod_{\mu=1}^{v-1}\hat{\mathbf{X}}[t-\mu]\bigg)\mathbf{X}[t] \bigg\rVert^2_F\bigg\} \notag \\
		&=E\Bigg\{\Bigg\lVert \mathbf{H}[t]\mathbf{V}[t]-\left[\sum_{v=1}^{V}p_v\bigg(\mathbf{H}[t-v]\mathbf{V}[t-v]+\mathbf{N}[t-v]\bigg)\prod_{\mu=1}^{v-1}\hat{\mathbf{X}}[t-\mu]\right]\mathbf{X}[t] \Bigg\rVert^2_F\Bigg\} \notag \\
		&=E\Bigg\{\Bigg\lVert \mathbf{H}[t]\mathbf{V}[t]-\left[\sum_{v=1}^{V}p_v\bigg(\mathbf{H}[t-v]\mathbf{V}[t-v]\prod_{\mu=1}^{v-1}\hat{\mathbf{X}}[t-\mu]+\mathbf{N}[t-v]\prod_{\mu=1}^{v-1}\hat{\mathbf{X}}[t-\mu]\bigg)\right]\mathbf{X}[t] \Bigg\rVert^2_F\Bigg\} \notag \\
		&=E\Bigg\{\Bigg\lVert \mathbf{H}[t]\mathbf{V}[t]-\left[\sum_{v=1}^{V}p_v\bigg(\mathbf{H}[t-v]\mathbf{V}[t-1]+\mathbf{N}[t-v]\prod_{\mu=1}^{v-1}\hat{\mathbf{X}}[t-\mu]\bigg)\right]\mathbf{X}[t] \Bigg\rVert^2_F\Bigg\} \notag \\
		&=E\Bigg\{\Bigg\lVert \mathbf{H}[t]\mathbf{V}[t]-\left[\sum_{v=1}^{V}p_v\bigg(\mathbf{H}[t-v]\mathbf{V}[t-1]\mathbf{X}[t]+\mathbf{N}[t-v]\prod_{\mu=1}^{v-1}\hat{\mathbf{X}}[t-\mu]\mathbf{X}[t]\bigg)\right] \Bigg\rVert^2_F\Bigg\} \notag \\
		&=E\bigg\{\bigg\lVert \mathbf{H}[t]\mathbf{V}[t]-\sum_{v=1}^{V}p_v\bigg(\mathbf{H}[t-v]\mathbf{V}[t]+\mathbf{N}[t-v]\prod_{\mu=0}^{v-1}\mathbf{X}[t-\mu]\bigg) \bigg\rVert^2_F\bigg\} \notag \\
		&=E\bigg\{\bigg\lVert \mathbf{H}[t]\mathbf{V}[t]-\sum_{v=1}^{V}p_v\bigg(\mathbf{H}[t-v]\mathbf{V}[t]+\mathbf{N}[t-v]\mathbf{V}^{-1}[t-v]\mathbf{V}[t]\bigg) \bigg\rVert^2_F\bigg\} \notag \\
		&=E\bigg\{\bigg\lVert \mathbf{H}[t]\mathbf{V}[t]-\sum_{v=1}^{V}p_v\bigg(\mathbf{H}[t-v]+\mathbf{N}[t-v]\mathbf{V}^{-1}[t-v]\bigg)\mathbf{V}[t] \bigg\rVert^2_F\bigg\} \notag\\
		&=E\bigg\{\bigg\lVert \bigg(\mathbf{H}[t]-\sum_{v=1}^{V}p_v\bigg(\mathbf{H}[t-v]+\mathbf{N}[t-v]\mathbf{V}^{-1}[t-v]\bigg)\bigg)\bigg(\mathbf{V}[t]\bigg)\bigg\rVert^2_F\bigg\}\notag\\
		&=E\bigg\{\bigg\lVert \mathbf{H}[t]-\sum_{v=1}^{V}p_v\bigg(\mathbf{H}[t-v]+\mathbf{N}[t-v]\mathbf{V}^{H}[t-v]\bigg) \bigg\rVert^2_F\bigg\}. \label{eq:msesimple}
	\end{align}
	We have $\mathbf{V}[t]=\mathbf{I}\mathbf{X}[1]\mathbf{X}[2]\cdots\mathbf{X}[t]$ from (\ref{eq:diffencode}). The matrix $\mathbf{X}[t]$ is unitary because it is a product of the unitary matrix $\mathbf{Z}[t]$ and a diagonal matrix $\mathbf{S}[t]$ with diagonal elements being $M$-PSK symbols. Hence $\mathbf{Z}^H[t]=\mathbf{Z}^{-1}[t]$ and $\mathbf{S}^{H}[t]\mathbf{S}[t]=\mathbf{S}[t]\mathbf{S}^{H}[t]=\mathbf{I}$, $\mathbf{X}^H[t]\mathbf{X}[t]=\mathbf{S}^{H}[t]\mathbf{Z}^H[t]\mathbf{Z}[t]\mathbf{S}[t]=\mathbf{I}=\mathbf{X}[t]\mathbf{X}^H[t]$. We have $\mathbf{S}^{H}[t]\mathbf{S}[t]=\mathbf{S}[t]\mathbf{S}^{H}[t]=\mathbf{I}$  because the product of any $M$-PSK symbol and its conjugate is 1. Furthermore, the products of unitary matrices are  unitary, hence $\mathbf{V}^H[t]=\mathbf{V}^{-1}[t]$, and $\mathbf{V}^H[t]\mathbf{V}[t]=\mathbf{V}[t]\mathbf{V}^H[t]=\mathbf{I}$.  The last step follows from the unitary invariance of the Frobenius matrix norm.

	The prediction coefficients $p_v$ that minimize $\sigma^2_{\rm mse}$  are formally  given by
	\begin{align}
		p_v = \arg \min_{p_v} E\bigg\{\bigg\lVert \mathbf{H}[t]-\sum_{v=1}^{V}p_v\bigg(\mathbf{H}[t-v]+\mathbf{N}[t-v]\mathbf{V}^{H}[t-v]\bigg) \bigg\rVert^2_F\bigg\}
\label{eq:minimization}	
      \end{align}
	The objective function can be expressed as
	\begin{align}
		L=E\bigg\{\bigg\lVert \mathbf{H}[t]-\sum_{v=1}^{V}p_v\bigg(\mathbf{H}[t-v]+\mathbf{N}[t-v]\mathbf{V}^{H}[t-v]\bigg) \bigg\rVert^2_F\bigg\}.
	\end{align} 
Since $L$ is quadratic with respect to $p_v$, the optimal solution to the minimization problem in (\ref{eq:minimization}) is obtained when $\frac{\partial L}{\partial p_v}=0, v=1, ...mV$, where $p_v$ is  complex scalar. The complex derivatives $\frac{\partial L}{\partial p_v}$ are the components of the complex gradient with respect to $p_v, v=1, ... , V$. The derivative of a real-value function $f$ with respect to a complex variable $x$ (Wirtinger derivative) is defined as \cite{quadraoptimal1,quadraoptimal2}
	\begin{align}
		\frac{\partial f(x)}{\partial x} =\frac{1}{2}\Bigg(\frac{\partial f(x)}{\partial {\rm Re}(x)}-{\rm j}\frac{\partial f(x)}{\partial {\rm Im} (x)}\Bigg). \label{eq:realftocompx}
	\end{align}
	The derivative of a complex-valued function $g$ with respect to $x$ is defined by
	\begin{align}
		\frac{\partial g(x)}{\partial x} =& \frac{\partial {\rm Re}(g(x))}{\partial x}+{\rm j}\frac{\partial {\rm Im} (g(x))}{\partial x} \notag\\
		=& \frac{1}{2}\Bigg(\frac{\partial {\rm Re}(g(x))}{\partial {\rm Re}(x)}-{\rm j}\frac{\partial {\rm Re}(g(x))}{\partial {\rm Im} (x)}\Bigg)+\frac{{\rm j}}{2}\Bigg(\frac{\partial {\rm Im}(g(x))}{\partial {\rm Re}(x)}-{\rm j}\frac{\partial {\rm Im}(g(x))}{\partial {\rm Im} (x)}\Bigg). \label{eq:compgtocompx}
	\end{align}
	Some results from \cite{quadraoptimal2} are used in calculating the complex gradient: $\frac{\partial x}{\partial x}=1$, $\frac{\partial x}{\partial x^*}=0$, and $\frac{\partial (x\cdot x^*)}{x}=x^*$, since following (\ref{eq:compgtocompx}), and with $x=a+jb$ and $a,b$ being real we have 
	\begin{align}
		\frac{\partial (x\cdot x^*)}{\partial x}&=\frac{\partial (a+bj)(a-bj)}{\partial(a+bj)}=\frac{\partial (a^2+b^2)}{\partial(a+bj)}\notag\\
		&=\frac{1}{2}(\frac{\partial(a^2+b^2)}{\partial a}-{\rm j}\frac{\partial(a^2+b^2)}{\partial b}) 
		=\frac{1}{2}(2a-j2b)=a-bj
		=x^*.
	\end{align}
	If a complex-valued function $g$ does not depend on $x^*$ but only on $x$, then $\frac{\partial g}{\partial x^*}=0$ \cite{quadraoptimal2}.
	
	The complex gradient of $L$ with respect to a complex variable $p_v$ can be expressed as
	\begin{align}
		\frac{\partial L}{\partial p_v}=&\frac{\partial E\bigg\{\bigg\lVert \mathbf{H}[t]-\sum_{v=1}^{V}p_v\bigg(\mathbf{H}[t-v]+\mathbf{N}[t-v]\mathbf{V}^{H}[t-v]\bigg) \bigg\rVert^2_F\bigg\}}{\partial p_v} \notag\\
		=&\frac{\partial}{\partial p_v} E\Bigg\{{\rm Tr}\bigg( \mathbf{H}[t]-\sum_{v=1}^{V}p_v\bigg(\mathbf{H}[t-v]+\mathbf{N}[t-v]\mathbf{V}^{H}[t-v]\bigg)\bigg)\notag\\
		&\bigg( \mathbf{H}^H[t]-\sum_{v=1}^{V}p_v^*\bigg(\mathbf{H}^H[t-v]+\mathbf{V}[t-v]\mathbf{N}^H[t-v]\bigg)\Bigg\}\notag
		\\
		=&\frac{\partial}{\partial p_v} \Bigg\{ E\bigg({\rm Tr}\Big(\mathbf{H}[t]\mathbf{H}^H[t]\Big)\bigg) - E\bigg({\rm Tr}\Big(\sum_{v=1}^V p_v^*\Big(\mathbf{H}[t]\mathbf{H}^H[t-v]+\mathbf{H}[t]\mathbf{V}[t-v]\mathbf{N}^H[t-v]\Big)\Big)\bigg) \notag\\
		-& E\bigg({\rm Tr}\Big(\sum_{v=1}^V p_v\Big(\mathbf{H}[t-v]\mathbf{H}^H[t]+\mathbf{N}[t-v]\mathbf{V}^H[t-v]\mathbf{H}^H[t]\Big)\Big)\bigg) \notag\\
		+& E\bigg({\rm Tr}\Big(\sum_{v=1}^V p_{v}\cdot \sum_{v'=1}^V p_{v'}^*\Big(\mathbf{H}[t-v]\mathbf{H}^H[t-v']+\mathbf{H}[t-v]\mathbf{V}[t-v']\mathbf{N}^H[t-v']\notag\\
		+&\mathbf{N}[t-v]\mathbf{V}^H[t-v]\mathbf{H}^H[t-v']+ \mathbf{N}[t-v]\mathbf{V}^H[t-v]\mathbf{V}[t-v']\mathbf{N}^H[t-v']\Big)\Big)\bigg) \Bigg\} 
	\end{align}.
	We have
	\vspace*{-6mm}
	\begin{align}
		\frac{\partial}{\partial p_v}\Biggl\{E\bigg({\rm Tr}\Big(\mathbf{H}[t]\mathbf{H}^H[t]\Big)\bigg)\Biggr\}=&0,\\
		\frac{\partial}{\partial p_v}\Biggl\{E\bigg({\rm Tr}\Big(\sum_{v=1}^V p_v^*\Big(\mathbf{H}[t]\mathbf{H}^H[t-v]+\mathbf{H}[t]\mathbf{V}[t-v]&\mathbf{N}^H[t-v]\Big)\Big)\bigg)\Biggr\}=0,
	\end{align}
	since they do not dependent on $p_v$ and their derivative with respect to $p_v$ is zero.
	Hence, 
	\begin{align}
		\frac{\partial L}{\partial p_v}=&\frac{\partial }{\partial p_v}\Biggl\{- E\bigg({\rm Tr}\Big(\sum_{v=1}^V p_v\Big(\mathbf{H}[t-v]\mathbf{H}^H[t]+\mathbf{N}[t-v]\mathbf{V}^H[t-v]\mathbf{H}^H[t]\Big)\Big)\bigg)\notag\\
		+&E\bigg({\rm Tr}\Big(\sum_{v=1}^V p_{v}\cdot \sum_{v'=1}^V p_{v'}^*\Big(\mathbf{H}[t-v]\mathbf{H}^H[t-v']+\mathbf{H}[t-v]\mathbf{V}[t-v']\mathbf{N}^H[t-v']\notag\\
		+&\mathbf{N}[t-v]\mathbf{V}^H[t-v]\mathbf{H}^H[t-v']+ \mathbf{N}[t-v]\mathbf{V}^H[t-v]\mathbf{V}[t-v']\mathbf{N}^H[t-v']\Big)\Big)\bigg) \Bigg\} \notag\\
		=&- E\bigg({\rm Tr}\Big(\mathbf{H}[t-v]\mathbf{H}^H[t]+\mathbf{N}[t-v]\mathbf{V}^H[t-v]\mathbf{H}^H[t]\Big)\bigg)\notag\\
		+& E\bigg({\rm Tr}\Big(\sum_{v'=1}^V p_{v'}^*\Big(\mathbf{H}[t-v]\mathbf{H}^H[t-v']+\mathbf{H}[t-v]\mathbf{V}[t-v']\mathbf{N}^H[t-v']\notag\\
		+&\mathbf{N}[t-v]\mathbf{V}^H[t-v]\mathbf{H}^H[t-v']+ \mathbf{N}[t-v]\mathbf{V}^H[t-v]\mathbf{V}[t-v']\mathbf{N}^H[t-v']\Big)\Big)\bigg) \notag \\
		=&- E\bigg({\rm Tr}\Big(\mathbf{H}[t-v]\mathbf{H}^H[t]\Big)\bigg)-E\bigg({\rm Tr}\Big(\mathbf{N}[t-v]\mathbf{V}^H[t-v]\mathbf{H}^H[t]\Big)\bigg)\notag\\
		+&E\bigg({\rm Tr}\Big(\sum_{v'=1}^V p_{v'}^*\Big(\mathbf{H}[t-v]\mathbf{H}^H[t-v']\Big)\Big)\bigg)+E\bigg({\rm Tr}\Big(\sum_{v'=1}^V p_{v'}^*\Big(\mathbf{H}[t-v]\mathbf{V}[t-v']\mathbf{N}^H[t-v']\Big)\Big)\bigg)\notag\\
		+&E\bigg({\rm Tr}\Big(\sum_{v'=1}^V p_{v'}^*\Big(\mathbf{N}[t-v]\mathbf{V}^H[t-v]\mathbf{H}^H[t-v']\Big)\Big)\bigg)\notag\\
		+& E\bigg({\rm Tr}\Big(\sum_{v'=1}^V p_{v'}^*\Big(\mathbf{N}[t-v]\mathbf{V}^H[t-v]\mathbf{V}[t-v']\mathbf{N}^H[t-v']\Big)\Big)\bigg), \label{eq:simpleLtopv}
	\end{align}.
	\enlargethispage*{10mm}
	We have
	\begin{align}
		E\bigg({\rm Tr}\Big(\mathbf{N}[t-v]\mathbf{V}^H[t-v]\mathbf{H}^H[t]\Big)\bigg)&={\rm Tr}\bigg(E\Big(\mathbf{N}[t-v]\mathbf{V}^H[t-v]\mathbf{H}^H[t]\Big)\bigg)\\
		&={\rm Tr}\bigg(E\Big(\mathbf{N}[t-v]\Big)E\Big(\mathbf{V}^H[t-v]\Big)E\Big(\mathbf{H}^H[t]\Big)\bigg)
		=0,
	\end{align}
	since the  matrix $\mathbf{H}[t]$, the noise matrix $\mathbf{N}[t]$, and the transmitted signal $\mathbf{V}[t]$ are mutually independent, and the fading process and the noise are also spatially uncorrelated. In addition, 
	\begin{align}
		E\Big(\mathbf{N}[t]\Big)=\left[\begin{matrix}
			E(n[t]_{11}) & E(n[t]_{12}) & \cdots & E(n[t]_{1K}) \\
			E(n[t]_{21}) & E(n[t]_{22}) & \cdots & E(n[t]_{2K}) \\
			\vdots & \vdots & \ddots & \vdots\\
			E(n[t]_{N_r1}) & E(n[t]_{N_r2}) & \cdots & E(n[t]_{N_rK})
		\end{matrix}\right]=\left[\begin{matrix}
			0 & 0 & \cdots & 0\\
			0 & 0 & \cdots & 0\\
			\vdots & \vdots & \ddots & \vdots\\
			0 & 0 & \cdots & 0\\
		\end{matrix}\right].
	\end{align} Similarly, we can show
	\begin{align}
		&E\bigg({\rm Tr}\Big(\sum_{v'=1}^V p_{v'}^*\Big(\mathbf{H}[t-v]\mathbf{V}[t-v']\mathbf{N}^H[t-v']\Big)\Big)\bigg)=0,\\
		&E\bigg({\rm Tr}\Big(\sum_{v'=1}^V p_{v'}^*\Big(\mathbf{N}[t-v]\mathbf{V}^H[t-v]\mathbf{H}^H[t-v']\Big)\Big)\bigg)=0.
	\end{align}
	Finally, (\ref{eq:simpleLtopv}) becomes
	\begin{align}
		\frac{\partial L}{\partial p_v}=& - E\bigg({\rm Tr}\Big(\mathbf{H}[t-v]\mathbf{H}^H[t]\Big)\bigg) + E\bigg({\rm Tr}\Big(\sum_{v'=1}^V p_{v'}^*\Big(\mathbf{H}[t-v]\mathbf{H}^H[t-v']\Big)\Big)\bigg)\label{eq:simpleLtopv12}\\
		+& E\bigg({\rm Tr}\Big(\sum_{v'=1}^V p_{v'}^*\Big(\mathbf{N}[t-v]\mathbf{V}^H[t-v]\mathbf{V}[t-v']\mathbf{N}^H[t-v']\Big)\Big)\bigg)          \label{eq:simpleLtopv3}.
	\end{align}
	
	To evaluate the first term in (\ref{eq:simpleLtopv12}) we  write the resulting matrix in component form
	\begin{align}
		\mathbf{H}[t-v]\mathbf{H}^H[t] =&\left[\begin{matrix}
			h[t-v]_{11} & h[t-v]_{12} & \cdots & h[t-v]_{1K}\\
			h[t-v]_{21} & h[t-v]_{22} & \cdots & h[t-v]_{2K}\\
			\vdots & \vdots & \ddots & \vdots\\
			h[t-v]_{N_r1} & h[t-v]_{N_r2} & \cdots & h[t-v]_{N_rK}
		\end{matrix}\right]\left[\begin{matrix}
			h^*[t]_{11} & h^*[t]_{21} & \cdots & h^*[t]_{N_r1}\\
			h^*[t]_{12} & h^*[t]_{22} & \cdots & h^*[t]_{N_r2}\\
			\vdots & \vdots & \ddots & \vdots\\
			h^*[t]_{1K} & h^*[t]_{2K} & \cdots & h^*[t]_{N_rK}
		\end{matrix}\right]\notag\\
		=&\left[\begin{matrix}
			\sum_{i=1}^{K}h[t-v]_{1i}h^*[t]_{1i} & \sum_{i=1}^{K}h[t-v]_{1i}h^*[t]_{2i} & \cdots & \sum_{i=1}^{K}h[t-v]_{1i}h^*[t]_{N_ri}\\
			\sum_{i=1}^{K}h[t-v]_{2i}h^*[t]_{1i} & \sum_{i=1}^{K}h[t-v]_{2i}h^*[t]_{2i} & \cdots & \sum_{i=1}^{K}h[t-v]_{2i}h^*[t]_{N_ri}\\
			\vdots & \vdots & \ddots & \vdots\\
			\sum_{i=1}^{K}h[t-v]_{N_ri}h^*[t]_{1i} & \sum_{i=1}^{K}h[t-v]_{N_ri}h^*[t]_{2i} & \cdots & \sum_{i=1}^{K}h[t-v]_{N_ri}h^*[t]_{N_ri}
		\end{matrix}\right].
	\end{align}
	Since the fading processes are spatially uncorrelated, we have
	\begin{align}
		E\bigg({\rm Tr}\Big[&\mathbf{H}[t-v]\mathbf{H}^H[t]\Big]\bigg)\notag\\
		=&{\rm Tr}\Bigg(\left[\begin{matrix}
			\sum_{i=1}^{K}E(h[t-v]_{1i}h^*[t]_{1i}) & 0 & \cdots & 0\\ 0 & \sum_{i=1}^{K}E(h[t-v]_{2i}h^*[t]_{2i}) & \cdots & 0\\
			\vdots & \vdots & \ddots & \vdots\\
			0 & 0 & 0 & \sum_{i=1}^{K}E(h[t-v]_{N_ri}h^*[t]_{N_ri})
		\end{matrix}\right]\Bigg) \notag\\
		=&{\rm Tr}\Bigg(\left[\begin{matrix}
			KJ_0(2\pi f_DT_sv) & 0 & \cdots & 0\\ 0 & KJ_0(2\pi f_DT_sv) & \cdots & 0\\
			\vdots & \vdots & \ddots & \vdots\\
			0 & 0 & 0 & KJ_0(2\pi f_DT_sv)
		\end{matrix}\right]\Bigg) \notag\\
		=&N_rKJ_0(2\pi f_DT_sv), \label{eq:EHH}
	\end{align}
	
	Similarly, the second term in (\ref{eq:simpleLtopv12}) can be  expressed as
	\begin{align}
		\sum_{v'=1}^V p_{v'}^*E\bigg({\rm Tr}\Big[\mathbf{H}[t-v]\mathbf{H}^H[t-v']\Big]\bigg) = \sum_{v'=1}^V p_{v'}^*N_rKJ_0(2\pi f_DT_s(v'-v)). \label{eq:EHH2}
	\end{align}
	
	Then, to find the third term, we first derive
	\begin{align}
		&\mathbf{V}^H[t-v]\mathbf{V}[t-v']\notag\\
		=&\left[\begin{matrix}
			v^*[t-v]_{11} & v^*[t-v]_{21} & \cdots & v^*[t-v]_{K1}\\
			v^*[t-v]_{12} & v^*[t-v]_{22} & \cdots & v^*[t-v]_{K2}\\
			\vdots & \vdots & \ddots & \vdots\\
			v^*[t-v]_{1K} & v^*[t-v]_{2K} & \cdots & v^*[t-v]_{KK}
		\end{matrix}\right]\left[\begin{matrix}
			v[t-v']_{11} & v[t-v']_{12} & \cdots & v[t-v']_{1K}\\
			v[t-v']_{21} & v[t-v']_{22} & \cdots & v[t-v']_{2K}\\
			\vdots & \vdots & \ddots & \vdots\\
			v[t-v']_{K1} & v[t-v']_{K2} & \cdots & v[t-v']_{KK}
		\end{matrix}\right]\notag\\
		=&\left[\begin{matrix}
			\sum_{i=1}^{K}v^*[t-v]_{i1}v[t-v']_{i1} & \sum_{i=1}^{K}v^*[t-v]_{i1}v[t-v']_{i2} & \cdots & \sum_{i=1}^{K}v^*[t-v]_{i1}v[t-v']_{iK}\\
			\sum_{i=1}^{K}v^*[t-v]_{i2}v[t-v']_{i1} & \sum_{i=1}^{K}v^*[t-v]_{i2}v[t-v']_{i2} & \cdots & \sum_{i=1}^{K}v^*[t-v]_{i2}v[t-v']_{iK}\\
			\vdots & \vdots & \ddots & \vdots\\
			\sum_{i=1}^{K}v^*[t-v]_{iK}v[t-v']_{i1} & \sum_{i=1}^{K}v^*[t-v]_{iK}v[t-v']_{i2} & \cdots & \sum_{i=1}^{K}v^*[t-v]_{iK}v[t-v']_{iK}
		\end{matrix}\right].
	\end{align}
	Next, we have
	\begin{align}
		&\mathbf{N}[t-v]\mathbf{V}^H[t-v]\mathbf{V}[t-v']=\left[\begin{matrix}
			n[t-v]_{11} & n[t-v]_{12} & \cdots & n[t-v]_{1K}\\
			n[t-v]_{21} & n[t-v]_{22} & \cdots & n[t-v]_{2K}\\
			\vdots & \vdots & \ddots & \vdots\\
			n[t-v]_{N_r1} & n[t-v]_{N_r2} & \cdots & n[t-v]_{N_rK}
		\end{matrix}\right]\mathbf{V}^H[t-v]\mathbf{V}[t-v']\notag\\
		=&\left[\begin{matrix}
			\sum_{j=1}^{K}\sum_{i=1}^{K}n[t-v]_{1j}v^*[t-v]_{ij}v[t-v']_{i1} & \cdots &  \sum_{j=1}^{K}\sum_{i=1}^{K}n[t-v]_{1j}v^*[t-v]_{ij}v[t-v']_{iK}\\
			\vdots & \ddots & \vdots\\
			\sum_{j=1}^{K}\sum_{i=1}^{K}n[t-v]_{N_rj}v^*[t-v]_{ij}v[t-v']_{i1} & \cdots &  \sum_{j=1}^{K}\sum_{i=1}^{K}n[t-v]_{N_rj}v^*[t-v]_{ij}v[t-v']_{iK}\\
		\end{matrix}\right],
	\end{align}
	where the element at the $p$-th row and $q$-th column is $\sum_{j=1}^{K}\sum_{i=1}^{K}n[t-v]_{pj}v^*[t-v]_{ij}v[t-v']_{iq}, 1\leq p\leq N_r,1\leq q\leq K$.
	Finally, the resulting matrix of (\ref{eq:simpleLtopv3}) is 
	\begin{align}
		&\mathbf{N}[t-v]\mathbf{V}^H[t-v]\mathbf{V}[t-v']\mathbf{N}^H[t-v']\label{eq:NVVN}\\
		=&\mathbf{N}[t-v]\mathbf{V}^H[t-v]\mathbf{V}[t-v']\left[\begin{matrix}
			n^*[t-v']_{11} & n^*[t-v']_{21} & \cdots & n^*[t-v']_{N_r1}\\
			n^*[t-v']_{12} & n^*[t-v']_{22} & \cdots & n^*[t-v']_{N_r2}\\
			\vdots & \vdots & \ddots & \vdots\\
			n^*[t-v']_{1K} & n^*[t-v']_{2K} & \cdots & n^*[t-v']_{N_rK}
		\end{matrix}\right].
	\end{align}
	The element at the $p$-th row and $l$-th column in (\ref{eq:NVVN})  where $1\leq p,l\leq N_r$ is
	\begin{align}
		(\mathbf{N}[t-v]\mathbf{V}^H[t-v]\mathbf{V}[t-v']\mathbf{N}^H[t-v'])_{pl}=\sum_{q=1}^{K}\sum_{j=1}^{K}\sum_{i=1}^{K}n[t-v]_{pj}v^*[t-v]_{ij}v[t-v']_{iq}n^*[t-v']_{lq},
	\end{align} 
	
	Since the noise process is spatially uncorrelated, the expectation of (\ref{eq:NVVN}) has nonzero values only on its diagonal, and $E(n[t-v]_{pj}n^*[t-v']_{pq})$ is nonzero if and only if $j=q$ and $v=v'$. Thus, we can express the expectation of the $p$-th diagonal element in (\ref{eq:NVVN}) as
	\begin{align}
		&\!\!\!\!\!\!\!\!\!\!\!\!\!E((\mathbf{N}[t-v]\mathbf{V}^H[t-v]\mathbf{V}[t-v']\mathbf{N}^H[t-v'])_{pp})\!=\!E\Big(\sum_{q=1}^{K}\sum_{j=1}^{K}\sum_{i=1}^{K}n[t-v]_{pj}v^*[t-v]_{ij}v[t-v']_{iq}n^*[t-v']_{pq}\Big)\notag\\
		&\!\!\!\!\!\!\!\!\!\!\!\!\!\!\!\!=\!E(\sum_{j=1}^{K}\sum_{i=1}^{K}n[t-v]_{pj}v^*[t-v]_{ij}v[t-v']_{ij}n^*[t-v']_{pj})
		=\sum_{j=1}^{K}\sum_{i=1}^{K}E(n[t-v]_{pj}v^*[t-v]_{ij}v[t-v']_{ij}n^*[t-v']_{pj}).\label{eq:simpleNVVN}
	\end{align}
	
	When $v\neq v'$, it is known that the noise processes at different time indices are independent of each other as well as independent of the transmitted signal. Therefore, the expectation of the product of these independent random process is
	\begin{align}
		\!\!\!\!\!E(n[t-v]_{pj}v^*[t-v]_{ij}v[t-v']_{ij}n^*[t-v']_{pj})=&E(n[t-v]_{pj})E(v^*[t-v]_{ij}v[t-v']_{ij})E(n^*[t-v']_{pj})
		=0,
	\end{align}
	since the noise process has zero mean. Then, when $v=v'$, (\ref{eq:simpleNVVN}) can be written by
	\begin{align}
		\!\!\!\!&E((\mathbf{N}[t-v]\mathbf{V}^H[t-v]\mathbf{V}[t-v]\mathbf{N}^H[t-v])_{pp})=\sum_{j=1}^{K}\sum_{i=1}^{K}E(n[t-v]_{pj}v^*[t-v]_{ij}v[t-v]_{ij}n^*[t-v]_{pj})\notag\\
		=&\sum_{j=1}^{K}E(n[t-v]_{pj}n^*[t-v]_{pj}\sum_{i=1}^{K}v^*[t-v]_{ij}v[t-v]_{ij})
		=\sum_{j=1}^{K}E(n[t-v]_{pj}n^*[t-v]_{pj})
		=\sum_{j=1}^{K}\sigma^2
		=K\sigma^2,
	\end{align}
	where $\sum_{i=1}^{K}v^*[t]_{ij}v[t]_{ij}=1$ for a unitary matrix $\mathbf{V}[t]\in\mathbb{C}^{K\times K}$, because 
	\begin{align}
		\mathbf{I}=&\mathbf{V}^H[t]\mathbf{V}[t]=\left[\begin{matrix}
			v^*[t]_{11} & v^*[t]_{21} & \cdots & v^*[t]_{K1} \\
			v^*[t]_{12} & v^*[t]_{22} & \cdots & v^*[t]_{K2} \\
			\vdots & \vdots & \ddots & \vdots\\
			v^*[t]_{1K} & v^*[t]_{2K} & \cdots & v^*[t]_{KK} \\
		\end{matrix}\right]\left[\begin{matrix}
			v[t]_{11} & v[t]_{12} & \cdots & v[t]_{1K} \\
			v[t]_{21} & v[t]_{22} & \cdots & v[t]_{2K} \\
			\vdots & \vdots & \ddots & \vdots\\
			v[t]_{K1} & v[t]_{K2} & \cdots & v[t]_{KK} \\
		\end{matrix}\right]\notag\\
		=&\left[\begin{matrix}
			\sum_{i=1}^{K}v^*[t]_{i1}v[t]_{i1} & 0 & \cdots & 0 \\
			0 & \sum_{i=1}^{K}v^*[t]_{i2}v[t]_{i2} & \cdots & 0 \\
			\vdots & \vdots & \ddots & \vdots\\
			0 & 0 & \cdots & \sum_{i=1}^{K}v^*[t]_{iK}v[t]_{iK} \\
		\end{matrix}\right].
	\end{align}
	Thus, the general expression of (\ref{eq:simpleNVVN}) is
	\begin{align}
		E((\mathbf{N}[t-v]\mathbf{V}^H[t-v]\mathbf{V}[t-v']\mathbf{N}^H[t-v'])_{pp})=K\sigma^2\delta[v'-v],
	\end{align} 
	where $\delta[m]=1$ only if $m=0$ and zero otherwise. As a result, (\ref{eq:simpleLtopv3}) is 
	\begin{align}
		\!\!\!\!\!\!\!\!\!\!\!\!\!\!\sum_{v'=1}^V p_{v'}^*E\bigg({\rm Tr}\Big[\mathbf{N}[t-v]\mathbf{V}^H[t-v]\mathbf{V}[t-v']\mathbf{N}^H[t-v']\Big]\bigg) =& \sum_{v'=1}^V p_{v'}^*{\rm Tr}\Bigg(\left[\begin{matrix}
			K\sigma^2\delta[v'-v] & \cdots & 0\\ 
			\vdots & \ddots & \vdots\\
			0 & \cdots & K\sigma^2\delta[v'-v]
		\end{matrix}\right]\Bigg) \notag \\
		=& \sum_{v'=1}^V p_{v'}^*N_rK\sigma^2\delta[v'-v]. \label{eq:NVVNresult}
	\end{align}

	By setting $\frac{\partial L}{\partial p_v}=0$ in (\ref{eq:simpleLtopv12}) and (\ref{eq:simpleLtopv3}), we can get
	\begin{align}
		0=\frac{\partial L}{\partial p_v}\notag
		=& - E\bigg({\rm Tr}\Big(\mathbf{H}[t-v]\mathbf{H}^H[t]\Big)\bigg) + E\bigg({\rm Tr}\Big(\sum_{v'=1}^V p_{v'}^*\Big(\mathbf{H}[t-v]\mathbf{H}^H[t-v']\Big)\Big)\bigg) \notag\\
		+& E\bigg({\rm Tr}\Big(\sum_{v'=1}^V p_{v'}^*\Big(\mathbf{N}[t-v]\mathbf{V}^H[t-v]\mathbf{V}[t-v']\mathbf{N}^H[t-v']\Big)\Big)\bigg),
	\end{align}
	which results in
	\begin{align}
		E\bigg({\rm Tr}\Big(\sum_{v'=1}^V p_{v'}^*\Big(\mathbf{H}[t-v]\mathbf{H}^H[t-v']\Big)\Big)\bigg) + E\bigg({\rm Tr}\Big(\sum_{v'=1}^V p_{v'}^*\Big(\mathbf{N}[t-v]\mathbf{V}^H&[t-v]\mathbf{V}[t-v']\mathbf{N}^H[t-v']\Big)\Big)\bigg)\notag\\
		&= E\bigg({\rm Tr}\Big(\mathbf{H}[t-v]\mathbf{H}^H[t]\Big)\bigg).
	\end{align}
	Then, from (\ref{eq:EHH}), (\ref{eq:EHH2}), and (\ref{eq:NVVNresult}), we have
	\begin{align}	
		&\sum_{v'=1}^V p_{v'}^*\big(N_rKJ_0(2\pi f_DT_s(v'-v))\big)+\sum_{v'=1}^V p_{v'}^*\big(N_rK\sigma^2\delta[v'-v]\big)=N_rKJ_0(2\pi f_DT_sv)\notag\\
		&\sum_{v'=1}^V p_{v'}^*\big(J_0(2\pi f_DT_s(v'-v))+\sigma^2\delta[v'-v]\big)=J_0(2\pi f_DT_sv), 1\leq v \leq V. \label{eq:LeqR}
	\end{align}
	
	To solve (\ref{eq:LeqR}), we convert the equations into matrix form. We define a matrix $\mathbf{R}\in \mathbb{C}^{V\times V}$ and a vector $\mathbf{b}\in\mathbb{C}^{V\times 1}$ as
	\begin{align}
		\mathbf{R}&=\left[\begin{matrix}
			J_0(0)+\sigma^2 & J_0(2\pi f_DT_s) & \cdots & J_0(2\pi f_DT_s(V-1))\\
			J_0(-2\pi f_DT_s) & J_0(0)+\sigma^2 & \cdots & J_0(2\pi f_DT_s(V-2))\\
			\vdots & \vdots & \ddots & \vdots\\
			J_0(2\pi f_DT_s(1-V)) & J_0(2\pi f_DT_s(2-V)) & \cdots & J_0(0)+\sigma^2 
		\end{matrix}\right],\label{eq:R}\\
		\mathbf{b} &= \left[\begin{matrix}
			J_0(2\pi f_DT_s) & J_0(2\pi f_DT_s\cdot2) & \cdots & J_0(2\pi f_DT_sV)
		\end{matrix}\right]^T.\label{eq:b}
	\end{align}
	It is clear that the element at the $v$-th row and $v'$-th column of $\mathbf{R}$ is $\mathbf{R}_{vv'}=J_0(2\pi f_DT_s(v'-v))+K\sigma^2\delta[v'-v]$ and the $v$-th element in $\mathbf{b}$ is $\mathbf{b}_v=J_0(2\pi f_DT_sv)$.
	Thus,  (\ref{eq:LeqR}) becomes
	\vspace*{-3mm}
	\begin{align}
		\mathbf{R}\mathbf{p}=\mathbf{b}, \label{eq:Rpb}
	\end{align}
	\vspace*{-5mm}
	where $\mathbf{p}=\left[\begin{matrix}
		p_1 & p_2 & \cdots & p_V
	\end{matrix}\right]^H$, with solution
	\begin{align}
		\mathbf{p}=\mathbf{R}^{-1}\mathbf{b}. \label{eq:pRb}
	\end{align}
	
	To prove  (\ref{eq:Rpb}), The left hand side is first expressed ain component form as
	\begin{align}
		\mathbf{R}\mathbf{p}=&\left[\begin{matrix}
			J_0(2\pi f_DT_s(1-1))+\sigma^2 & J_0(2\pi f_DT_s(2-1)) & \cdots & J_0(2\pi f_DT_s(V-1))\\
			J_0(2\pi f_DT_s(1-2)) & J_0(2\pi f_DT_s(2-2))+\sigma^2 & \cdots & J_0(2\pi f_DT_s(V-2))\\
			\vdots & \vdots & \ddots & \vdots\\
			J_0(2\pi f_DT_s(1-V)) & J_0(2\pi f_DT_s(2-V)) & \cdots & J_0(2\pi f_DT_s(V-V))+\sigma^2 
		\end{matrix}\right]\left[\begin{matrix}
			p_1^* \\ p_2^* \\ \vdots \\ p_V^*
		\end{matrix}\right]\notag\\
		=& \left[\begin{matrix}
			\sum_{v'=1}^V p_{v'}^*\big(J_0(2\pi f_DT_s(v'-1))+\sigma^2\delta[v'-1]\big) \\
			\sum_{v'=1}^V p_{v'}^*\big(J_0(2\pi f_DT_s(v'-2))+\sigma^2\delta[v'-2]\big) \\
			\vdots \\
			\sum_{v'=1}^V p_{v'}^*\big(J_0(2\pi f_DT_s(v'-V))+\sigma^2\delta[v'-V]\big)
		\end{matrix}\right], \label{eq:Rp}
	\end{align}
	and the right hand side is $\mathbf{b}$ of (\ref{eq:b}). Therefore, we have 
	\begin{align}
		\left[\begin{matrix}
			\sum_{v'=1}^V p_{v'}^*\big(J_0(2\pi f_DT_s(v'-1))+\sigma^2\delta[v'-1]\big) \\
			\sum_{v'=1}^V p_{v'}^*\big(J_0(2\pi f_DT_s(v'-2))+\sigma^2\delta[v'-2]\big) \\
			\vdots \\
			\sum_{v'=1}^V p_{v'}^*\big(J_0(2\pi f_DT_s(v'-V))+\sigma^2\delta[v'-V]\big)
		\end{matrix}\right]=\left[\begin{matrix}
			J_0(2\pi f_DT_s\cdot1) \\ J_0(2\pi f_DT_s\cdot2) \\ \cdots \\ J_0(2\pi f_DT_s\cdot V)
		\end{matrix}\right],
	\end{align}
	which coorresponds to  (\ref{eq:LeqR})  for $v=1,\cdots,V$.
	{
		\bibliographystyle{ieeetr}
		\bibliography{references}		
	}

\end{document}